\documentclass[usenatbib]{mn2e}
\usepackage{epsfig,amsmath}
\newcommand{\se}{\sigma_8}
\newcommand{\eg}{{\it e.g.\ }}
\newcommand{\ie}{{\it i.e. }}

\newcommand{\lteq}{\mathrel{\raise.3ex\hbox{$<$\kern-.75em\lower1ex\hbox{$\sim$}}}}
\newcommand{\gteq}{\mathrel{\raise.3ex\hbox{$>$\kern-.75em\lower1ex\hbox{$\sim$}}}}
\newcommand{\hmpc}{$h^{-1}$Mpc}
\newcommand{\degree}{{\rm o}}

\begin{document}
 
\title[Detected fluctuations in SDSS LRG magnitudes]{Detected fluctuations in SDSS LRG magnitudes: Bulk flow signature or systematic?}
\author[Abate and Feldman]{Alexandra Abate$^{1}$ and Hume A. Feldman$^{2}$\\
$^1$Physics Department, University of Arizona, 1118 East 4th Street, Tucson, AZ 85721, USA\\
$^2$Department of Physics and Astronomy, University of Kansas, Lawrence, KS 66045, USA\\}
\maketitle

\begin{abstract}
In this paper we search for a signature of a large scale bulk flow by looking for fluctuations in the magnitudes of distant LRGs.  We take a sample of LRGs from the Sloan Digital Sky Survey with redshifts of $z>0.08$ over a contiguous area of sky.  Neighboring LRG magnitudes are averaged together to find the fluctuation in magnitudes as a function of R.A.. The result is a fluctuation of a few percent in flux across roughly 100 degrees. The source of this fluctuation could be from a large 
dipole motion with respect to the LRG sample, 
 or a systematic in our treatment of the data set, or the data set itself. A 
 dipole 
 model is fitted to the observed fluctuation, and the three flow parameters, its direction and magnitude: $\alpha_b$, $\delta_b$, $v_b$ are constrained.  We find that the flow direction is consistent with the direction found by other authors, with $\alpha_b\sim 180$, $\delta_b\sim-50$.  The flow magnitude however was found to be anomalously large with $v_b>4000$km/s.  The LRG angular selection function cannot be sufficiently taken into account in our analysis with the available data, and may be the source of either the anomalous magnitude of the flow signal, or possibly the entire fluctuation. However, the fluctuation indicates a flow direction very close to those found using other data sets and analyses. Further investigation with upcoming data is required to confirm this detection.

\end{abstract}

\section{Introduction}
\renewcommand{\thefootnote}{\fnsymbol{footnote}}
\setcounter{footnote}{1}
\footnotetext{E-mail: abate@email.arizona.edu, feldman@ku.edu}

Recent years have seen a rapid growth of cosmological observations and analyses that has brought forth the era of {\it precision cosmology} led by missions like the WMAP satellite \citep{wmap3,wmap5,wmap7}, the SDSS collaboration (see at http://www.sdss.org), DEEP (http://deep.berkeley.edu) and many others. As important and interesting as these measurements are, probing the large scale structure of the Universe using many diverse and independent measurements remains of utmost importance. As much as we have learned in the last decade, cosmology and even parameter estimation is not a closed subject \citep{BriLahOstSte03}. Not only do we not know the nature of dark matter, but even the density and amplitude of its initial perturbations (usually parametrized by $\Omega_m$ and $\se$ respectively) remain an open question.  The determination of the matter content is especially problematic \citep{FukPee04,sigma810}. 

Further, there are some indications that suggest that the $\Lambda$CDM cosmology with WMAP \citep{wmap7} central parameters may be problematic. Some examples are: 
Large-scale anomalies found in the maps of temperature anisotropies in the CMB \citep{SarHutCop10,CopHutSch10,WMAPanom10};
recent estimates of the large scale bulk flow by \citet{WatFelHud09,FelWatHud10,MacFelFer11,MaGorFel11} are inconsistent at the nearly 3$\sigma$ level with $\Lambda$CDM predictions;  
a recent estimate \citep{LeeKom10} of the occurrence of high-velocity merging systems such as the Bullet Cluster is unlikely at a $\sim6\sigma$ level; 
large excess of power in the statistical clustering of luminous red galaxies (LRG) in the photometric SDSS galaxy sample 
\citep{ThoAbdLah11};
evidence of larger than expected cross correlation between samples of galaxies and lensing of the CMB \citep{HoHirPad08,HirHoPad08};
brighter than expected type Ia Supernovae (SNIa) at High Redshift \citep{Kowalski08}; 
voids, especially smaller ones ($\sim10$ Mpc) are observed to be much emptier than predicted \citep{GotLokKly03}; 
the predicted shallow low concentration and density profiles of Cluster Haloes disagree with observations which indicate denser high concentration cluster haloes \citep{deBlok05,Gentile05};
 \citet{KovBenItz10} find a unique direction in the CMB sky determined by anomalous mean temperature ring profiles, also centered about the direction of the flow detected above.

The amplitude and growth of cosmological fluctuations on large scales are closely related to the CMB dipole, which reflects the bulk flow (BF) of the local group and can be used to test cosmological models. Large scale velocity surveys have been undertaken by various groups (\eg \citet{GioHayHer97,HudSmiLuc99,HudSmiLuc04,SFI1,SFI2,sfierr09}) and recent analyses by \cite{WatFelHud09,FelWatHud10} of the newest proper distance measurements show that virtually all velocity survey analyses show a consistent large-scale BF. It appears to have a magnitude $\gteq400$ km/s on scales of 100\hmpc, which disagrees with $\Lambda$CDM WMAP predictions at the $\sim3\sigma$ level. Analysis of the flow \citep{FelWatHud10} suggests that if it is due to a gravitational potential flow, then the sources (\ie over- and under-densities) must be on scales $\ \gteq 300$\hmpc. The direction of this flow is close to the Galactic disk, Galactic longitude $\sim295^\degree $ and longitude $\sim10^\degree $ with error of $\sim5^\degree$. Assuming that this is a potential flow, we expect to see over-densities in the flow direction and under-densities in the opposite direction. The responsible structures must be far away from us since the flow is very cold \citep{FelWatHud10}.

The motion detected in \citet{FelWatHud10} is not due to nearby sources, such as the Great Attractor (distance of $\sim40$\hmpc), but rather to sources at greater depths that have yet to be fully identified. The largest known mass concentration, the Shapley supercluster, does not seem to be massive enough to cause a flow of this magnitude \citep{Ray89}. Following \citet{tully08}, it is more likely that the flow arises both from various mass concentrations in the Galactic y-direction as well as under-dense regions in the opposite direction. Currently, there is no survey in existence that is deep enough to resolve the source(s) of the flow.  Nor is there any data that probes these scales ($>300$\hmpc) to see whether there are any large mass concentrations and voids in these directions.
\begin{figure}
\begin{tabular}{c}
\includegraphics[width=6.5cm]{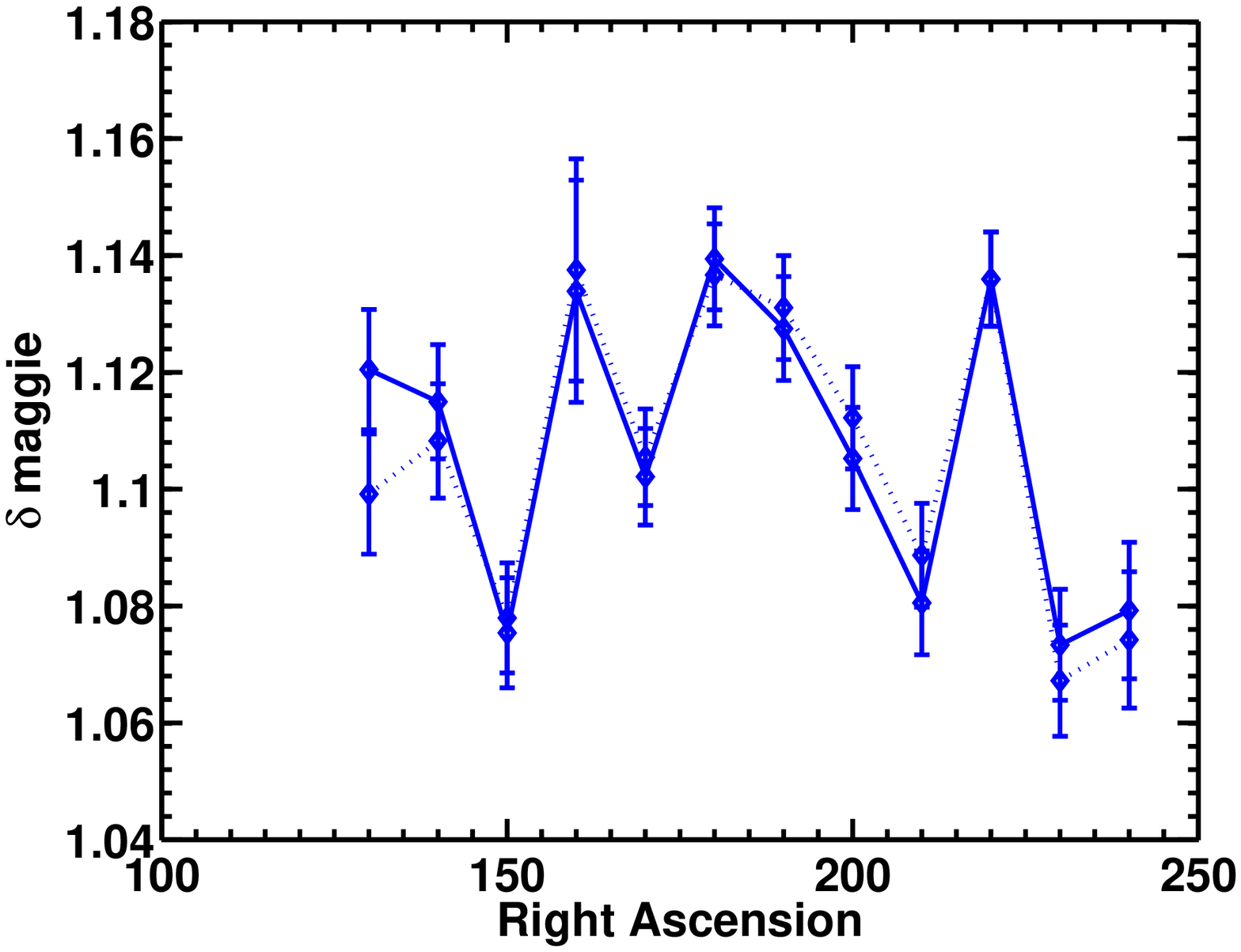} \\
\includegraphics[width=6.5cm]{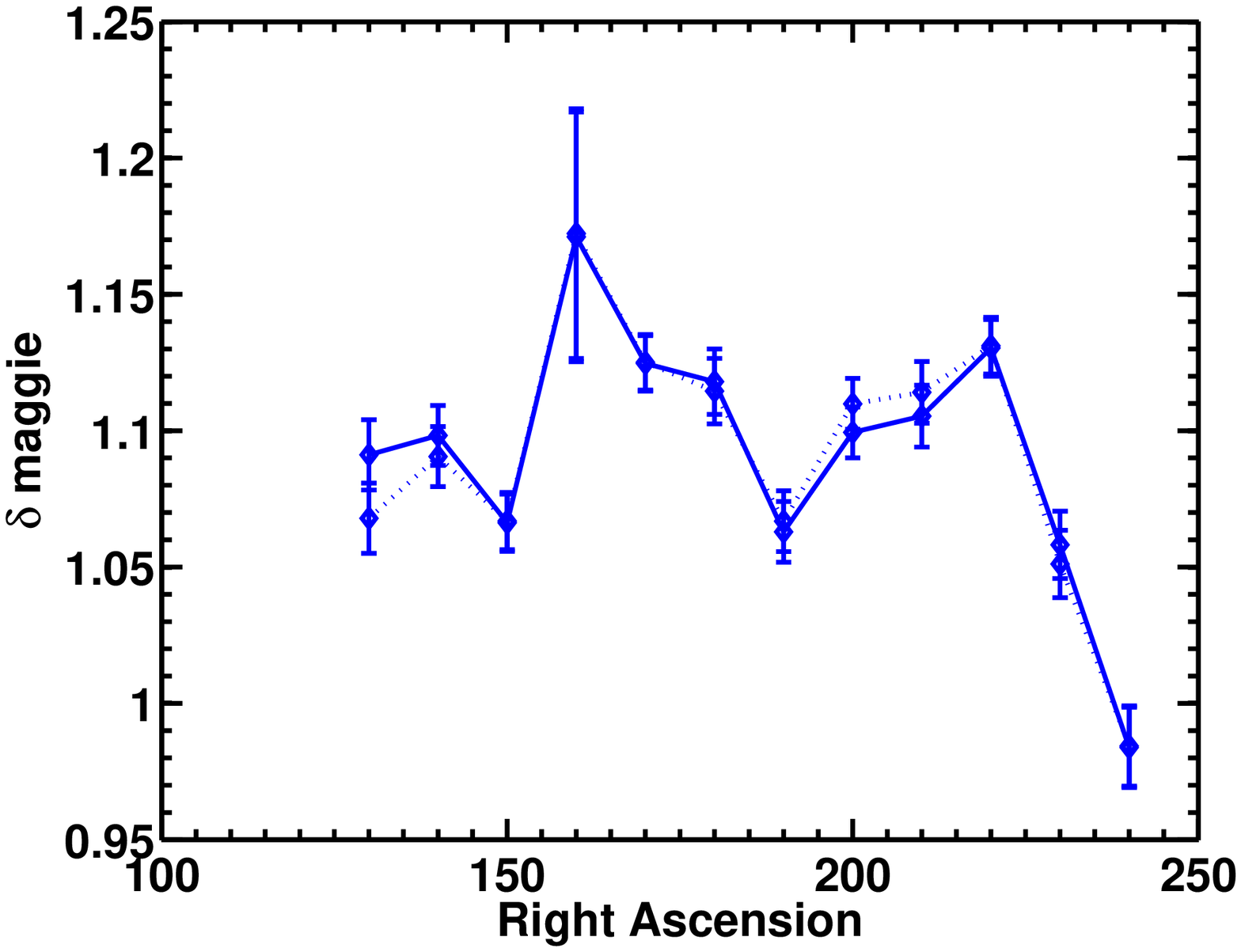} \\
\includegraphics[width=6.5cm]{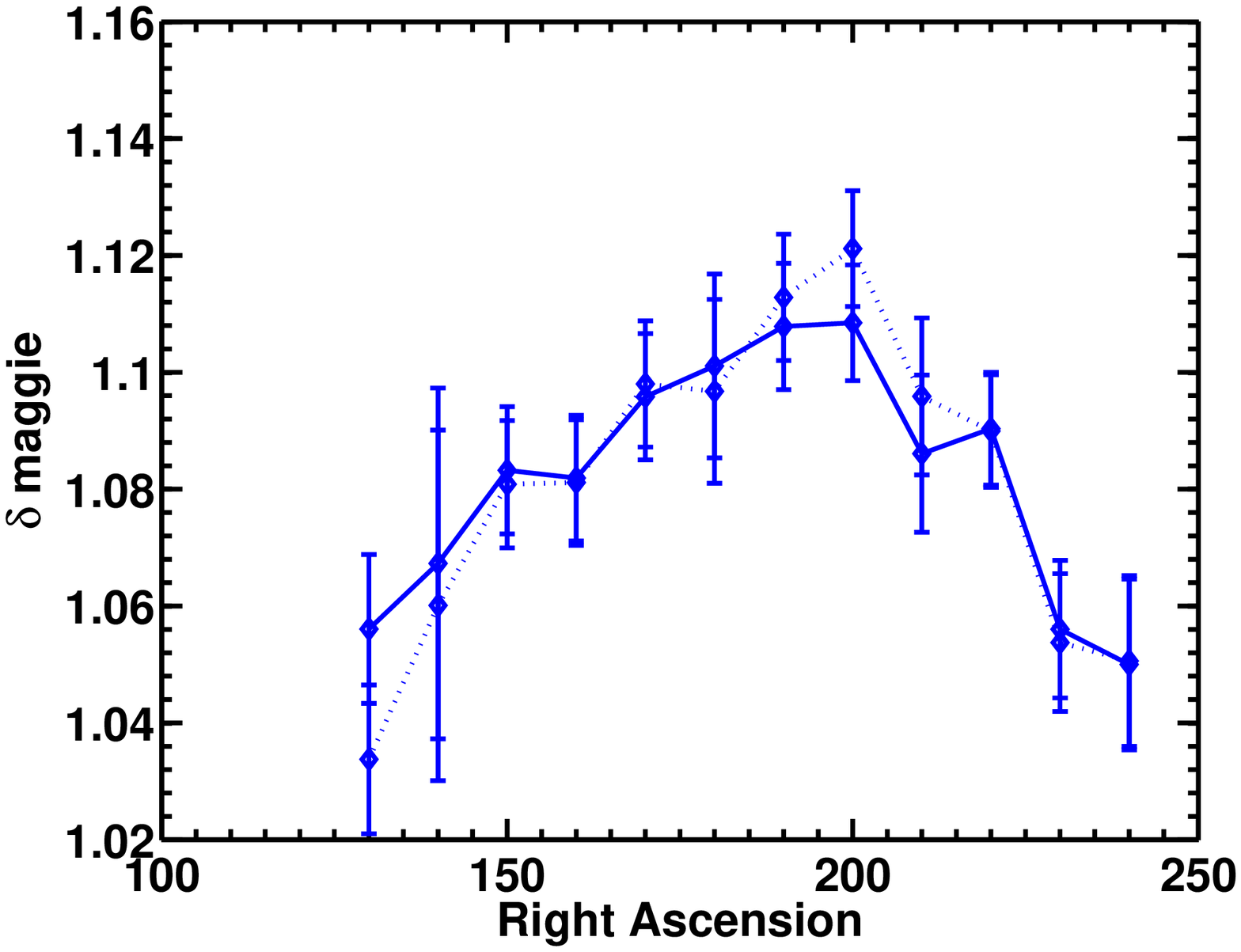} \\
\includegraphics[width=6.5cm]{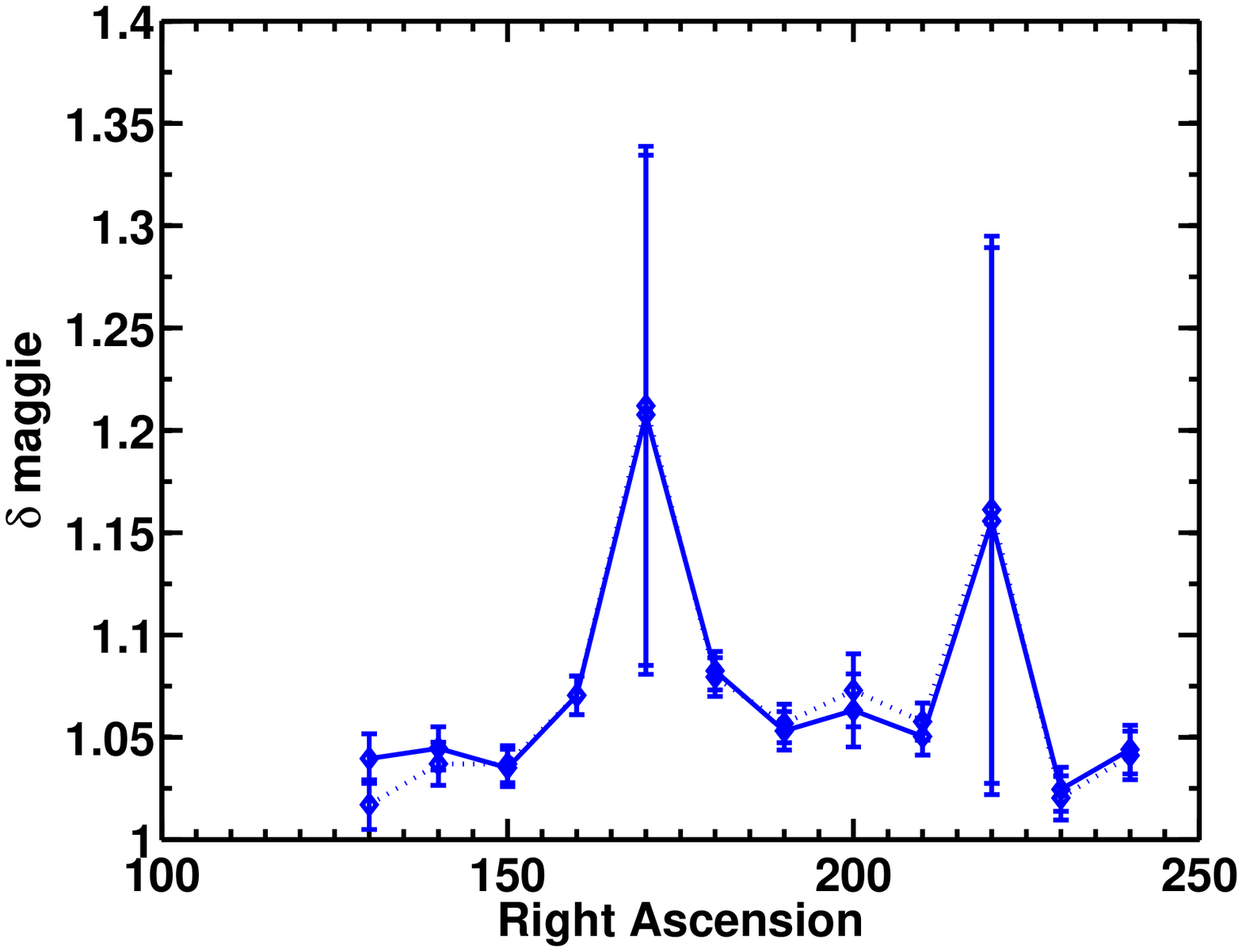} \\
\end{tabular}
\caption{
The magnitude fluctuation as a function of R.A.. The LRG $r$ band model magnitudes in bins of R.A. and redshift are averaged over all values of the declinations, the mean magnitude at each redshift is then subtracted.  The fluctuation is in units of ``maggies", see Eqs. \ref{eq:find},\ref{eq:find2}.  Each panel shows the different \textit{spectroscopic} redshift bins: top panel $z_c=0.12$, 2nd panel $z_c=0.20$, 3rd panel $z_c=0.28$, bottom panel $z_c=0.36$.  The solid lines are the Galactic extinction corrected magnitudes, and the dotted lines are the magnitudes with no extinction correction applied.
\label{fig:avmagind}
} 
\end{figure}

\citet{KasAtrKoc08,KasAtrKoc10} found a larger amplitude BF, coined the \textit{Dark Flow}, on scales out to 300 $h^{-1} {\rm Mpc}$ by using the kinematic Sunyaev-Zeldovich effect (kSZ) in the CMB to measure the bulk velocity of about 1200 X-ray clusters. Like \citet{WatFelHud09,FelWatHud10}, these papers claim that a BF at these depths, as determined by their studies, is difficult or impossible to explain within the framework of standard $\Lambda$CDM model of cosmology. Although these results utilize completely different methods to measure the BF and find similar results, kSZ and velocity surveys are both subject to problems of very large noise and systematics in the data.  In particular it should be noted that converting the dipole of the kSZ signal to a velocity is model dependent: assuming a different cluster radial profile can give a completely different result. However, other recent results \citep{DavNusMas11,NusDav11} using the SFI++ catalog of Tully-Fisher galaxies and also reconstructing the cosmological large-scale flows using the 2MASS (Two Micron All Sky Survey) redshift survey (2MRS) found flows roughly the same direction as mentioned above but with magnitude consistent with the WMAP $\Lambda$CDM model.

In this paper we approach the problem in a different manner and instead look for fluctuations of observed magnitudes across the sky as could be induced by a bulk motion centered on us.  We choose objects that we assume are distant enough to be considered at rest compared to the CMB.  Ideally these objects should be standard candles, therefore the best astrophysical candidate would be SNIa. Unfortunately there are not enough observations of SNIa, and their distribution across the sky is far from ideal for this investigation.  We choose instead to use luminous red galaxies (LRGs) which are suitable because they are assumed to form a single population of galaxies, which assembled at high-redshift and have been passively evolving since.  This assumption was investigated by \citet{WakNicEis06} who found that the LRG luminosity function (LF) is consistent with purely passive evolution and a merger-free history. Therefore to use the LRGs as a proxy for standard candles we average out the effect of the intrinsic variability of their luminosities by averaging together the magnitudes of neighboring LRGs. A recent study \citep{NusBraDav11} presented an alternate method for detecting cosmological bulk flows from redshift surveys, using the observed dimming or brightening of galaxies due to their peculiar motion, a similar technique to that which we present here.

The paper is organized as follows, in Section \ref{sec:data} we present our data set and how we determine the observed magnitude fluctuation, in Section \ref{sec:theory} we present the theoretical model for magnitude fluctuations induced by peculiar velocities. Section \ref{sec:fit} describes our method of analyzing the data, with the results presented in Sections \ref{sec:res} and \ref{sec:dir}, and discussion and conclusions of these results in Section \ref{sec:disc}. 

Throughout this paper we assume a concordance cosmology of: $H_0$=70km/s, $\Omega_m=0.3$, $\Omega_\Lambda=0.7$ and $w=-1$ unless otherwise stated.

\begin{figure}
\begin{tabular}{c}
\includegraphics[width=6.5cm]{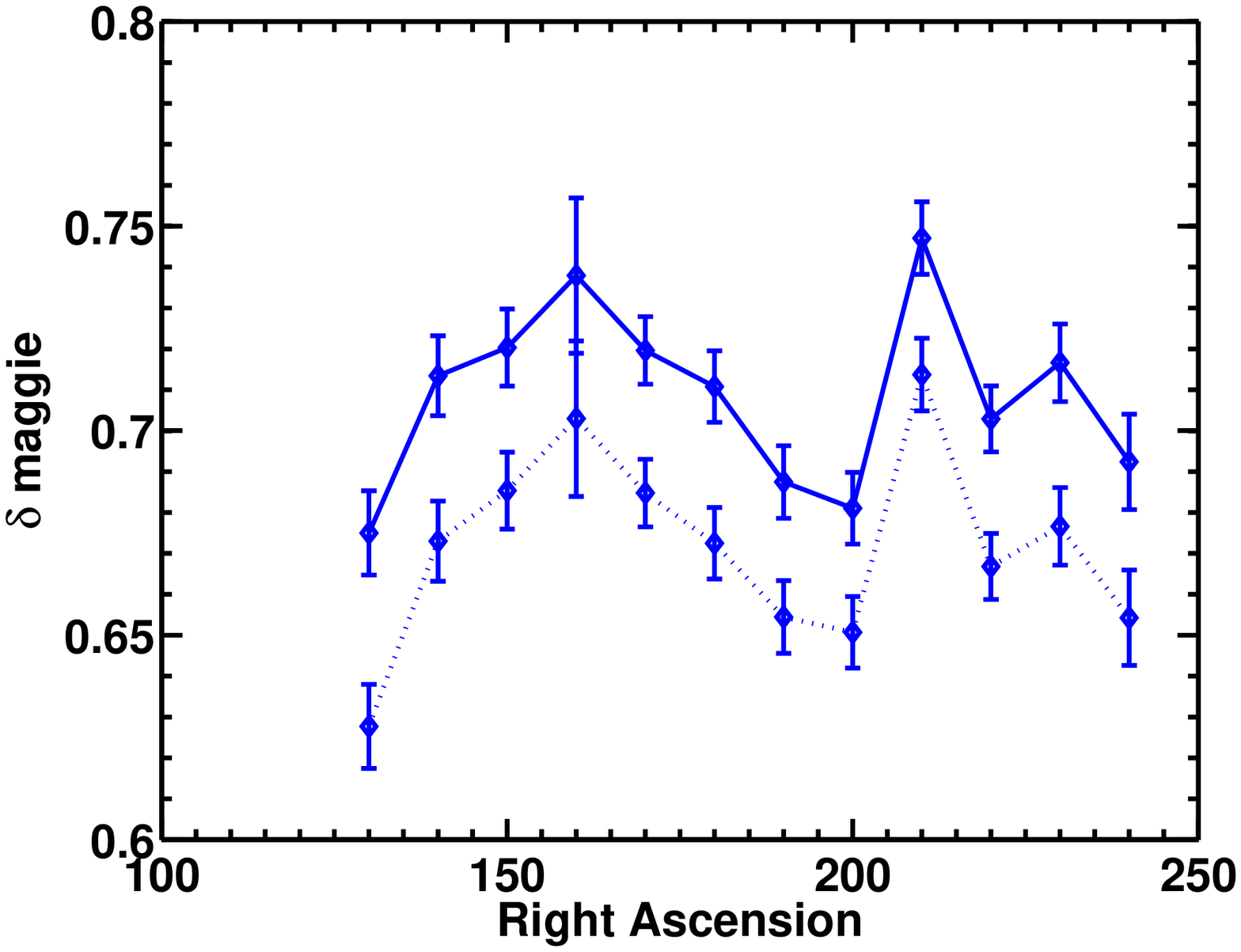} \\
\includegraphics[width=6.5cm]{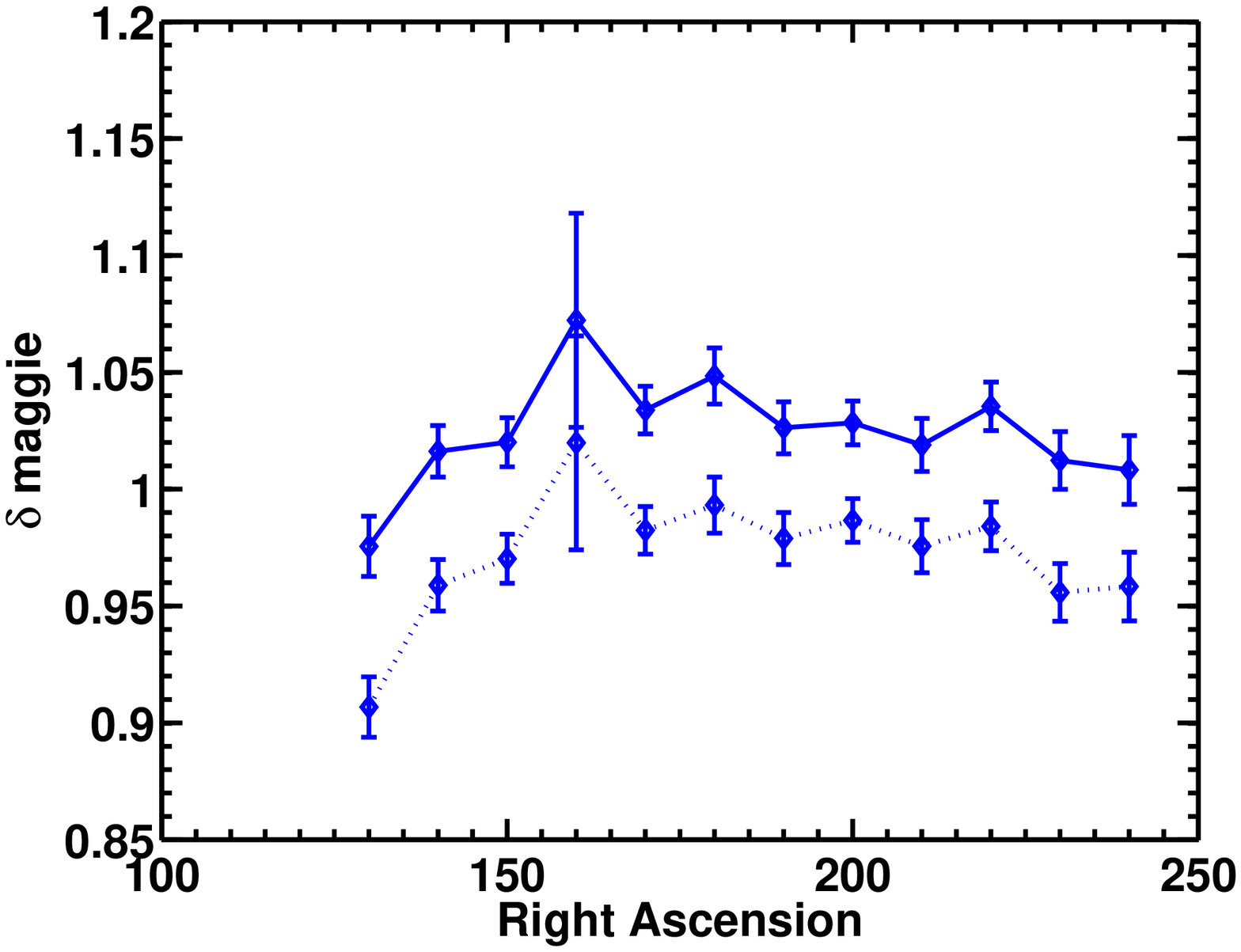} \\
\includegraphics[width=6.5cm]{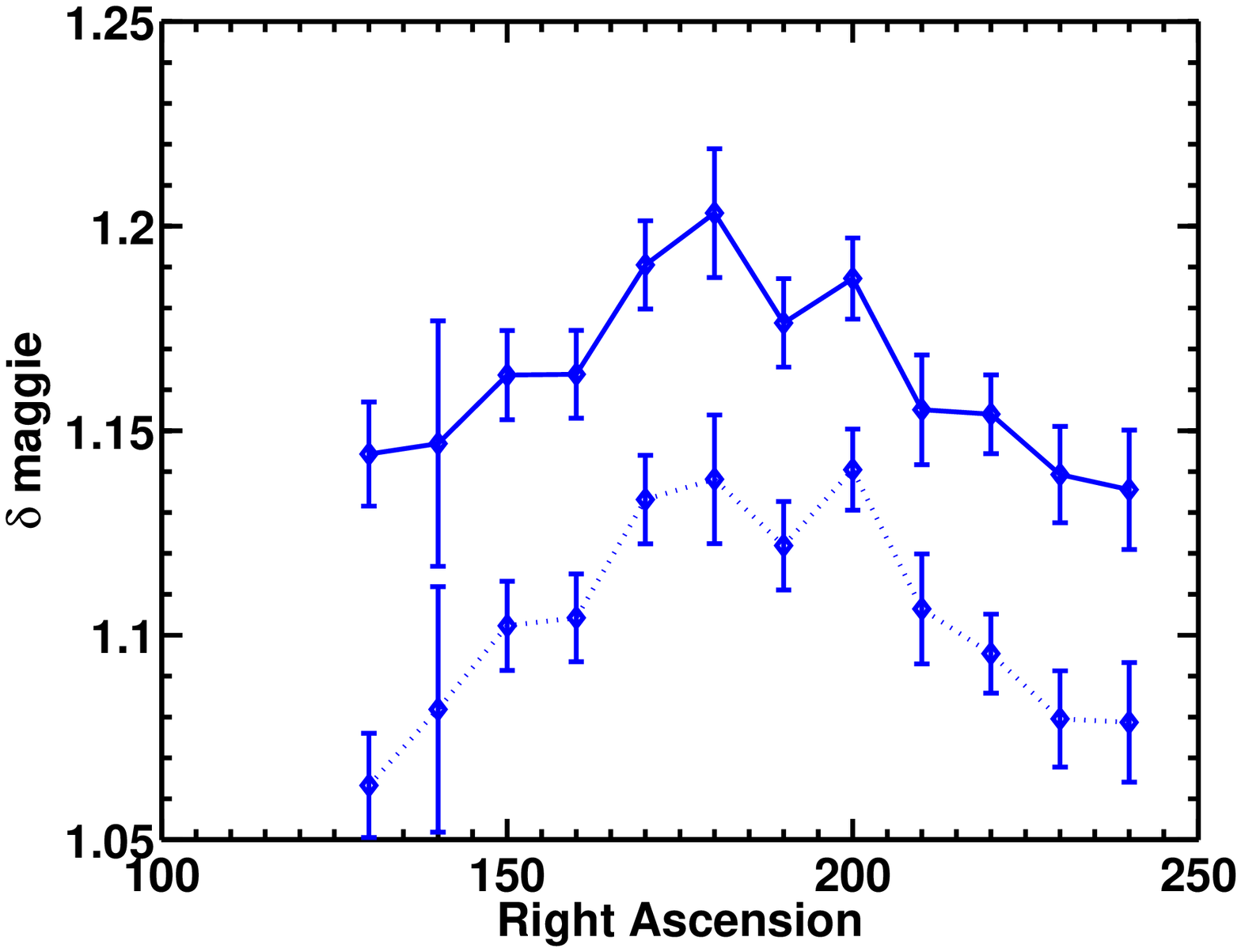} \\
\includegraphics[width=6.5cm]{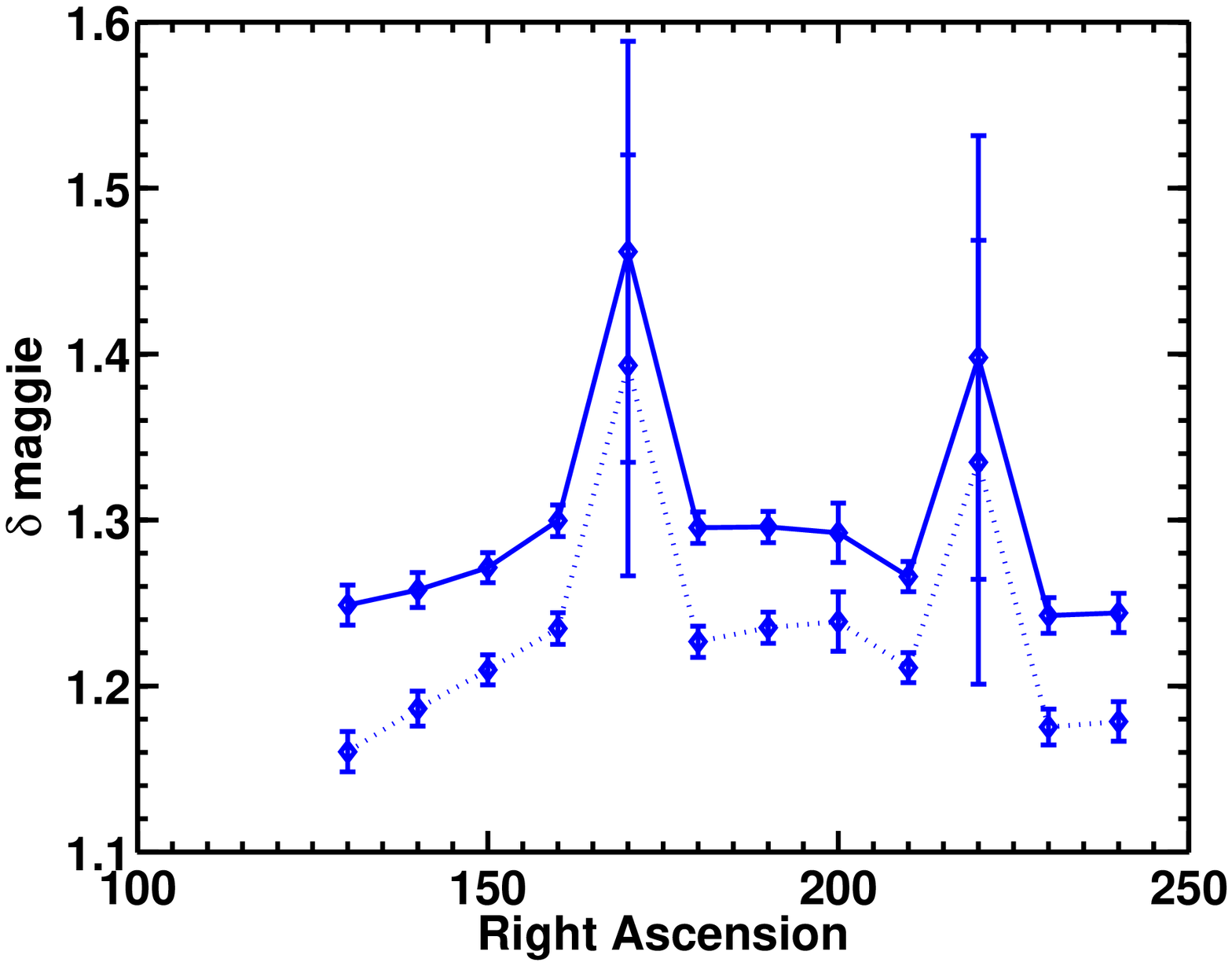} \\
\end{tabular}
\caption{
The ``model dependent" magnitude fluctuation as a function of R.A.. The difference between LRG $r$ band model magnitudes and the ``theoretical" catalog in bins of R.A. and redshift is averaged, over all values of the declinations.  The fluctuation is in units of ``maggies", see Eqs. \ref{eq:df},\ref{eq:df2}. Each panel shows a different \textit{spectroscopic} redshift bin: top panel $z_c=0.12$, 2nd panel $z_c=0.20$, 3rd panel $z_c=0.28$, bottom panel $z_c=0.36$.  The solid lines are the Galactic extinction corrected magnitudes, and the dotted lines are the magnitudes with no extinction correction applied.
\label{fig:avmag}
} 
\end{figure}

\section{The observed fluctuation}
\label{sec:data}
\begin{figure*}
\begin{tabular}{cc}
 \includegraphics[width=6.5cm]{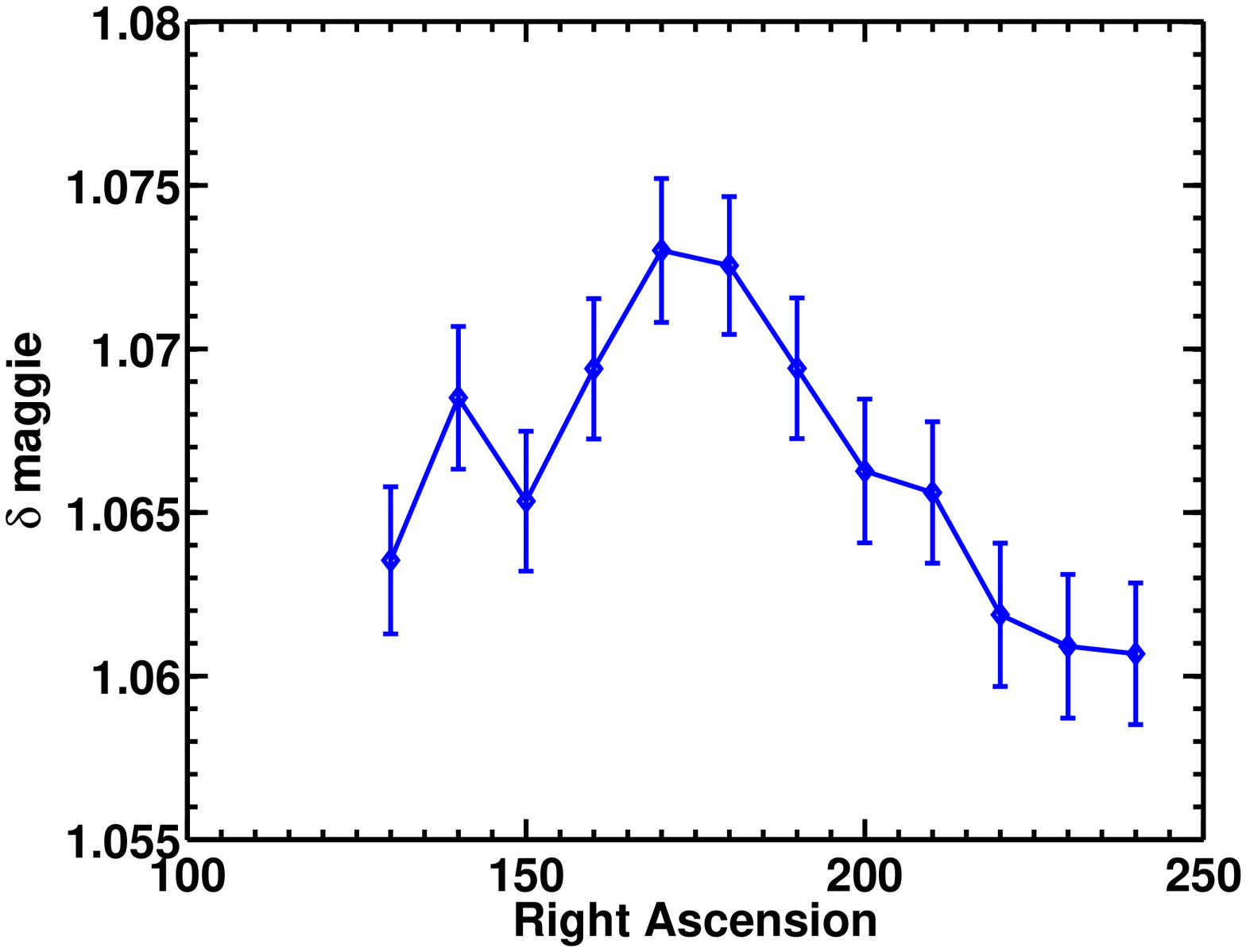} &  \includegraphics[width=6.5cm]{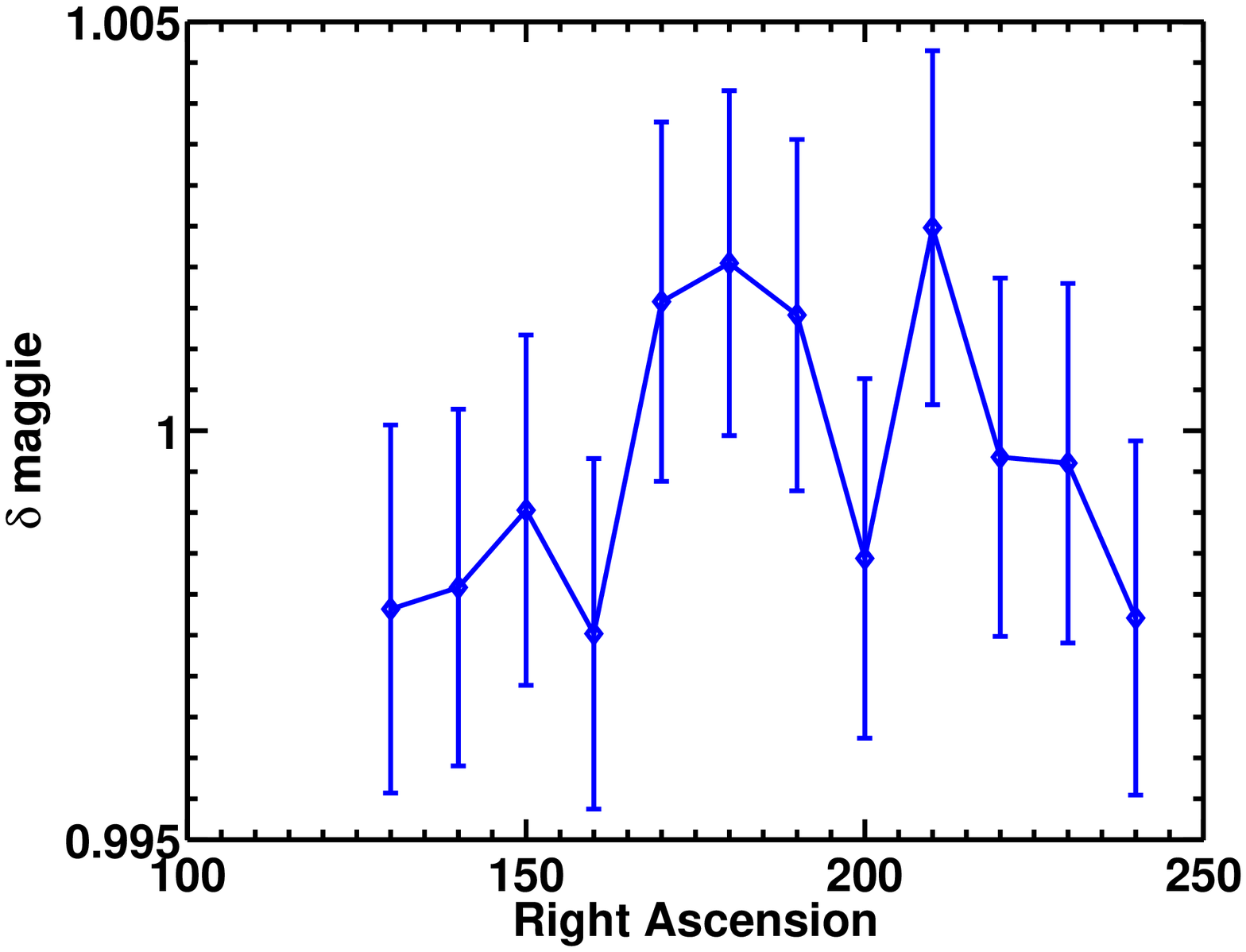} \\
 \includegraphics[width=6.5cm]{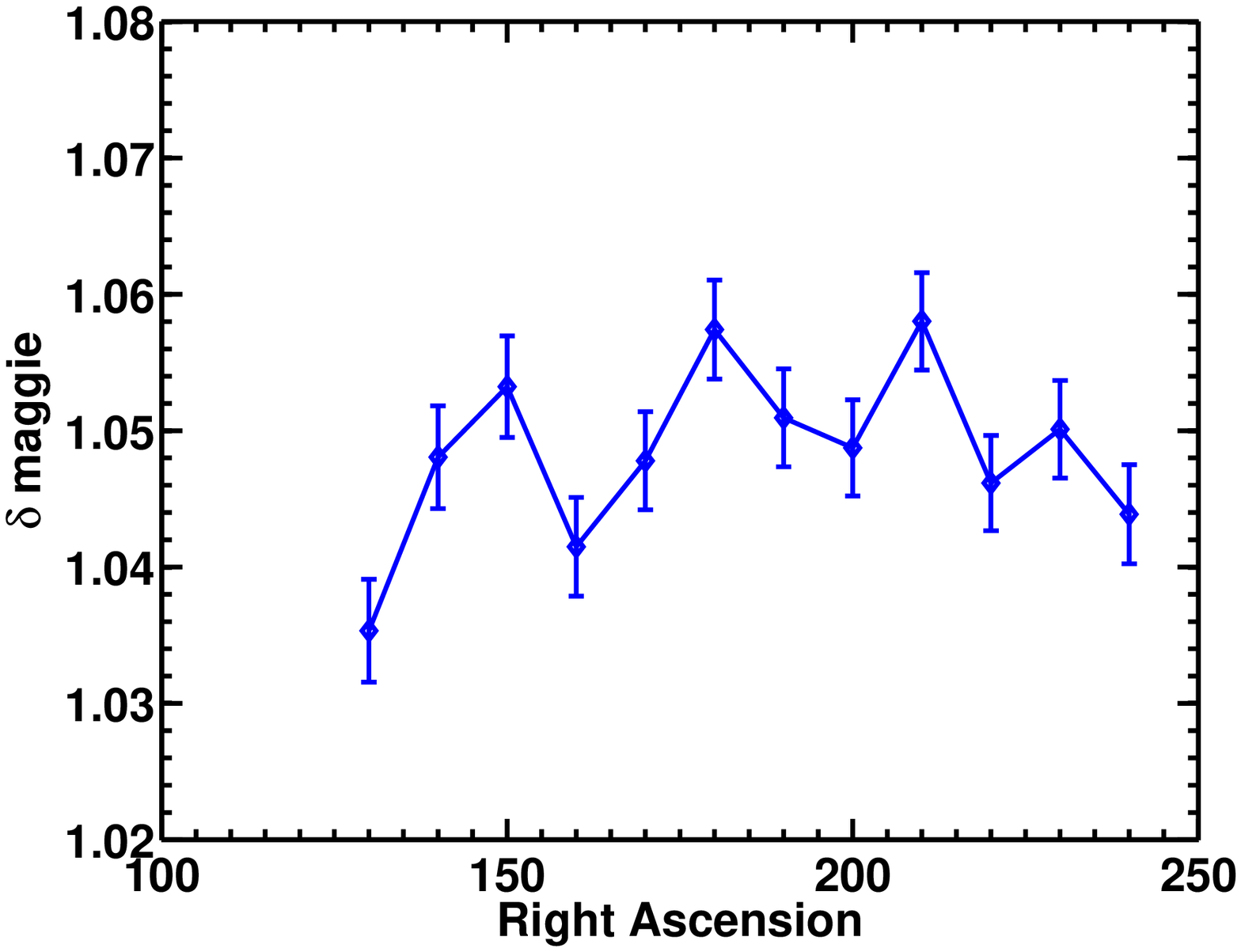} &  \includegraphics[width=6.5cm]{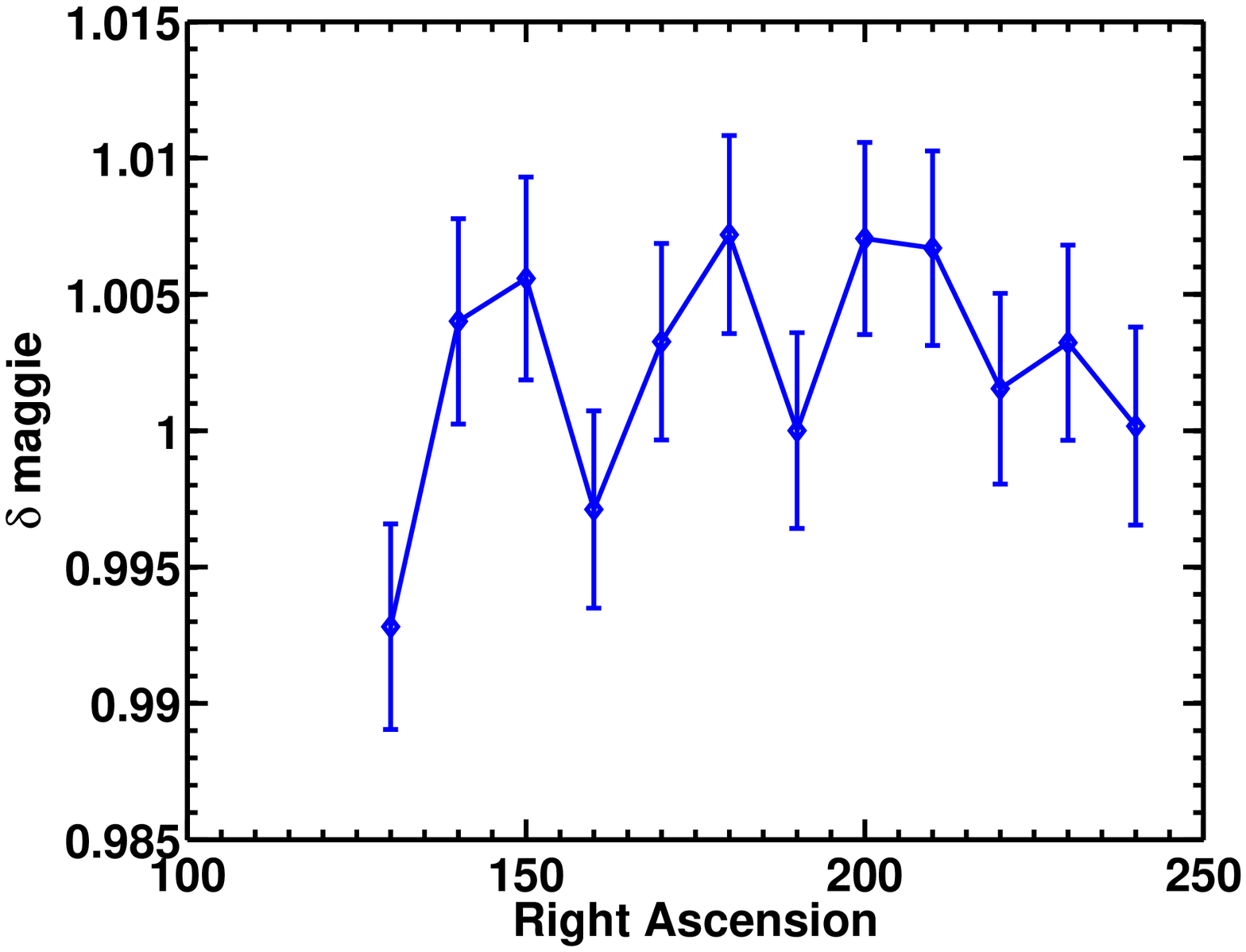} \\
\end{tabular}
\caption{
The magnitude fluctuation as a function of R.A. for the \textit{photometric} sample. Each row shows a different \textit{photometric} redshift bin: top row $z_c=0.475$, bottom row $z_c=0.625$.  The two columns correspond to the two different ways of computing the fluctuation: left column is the fluctuation described by Eq.~\ref{eq:find} (as in Fig.~\ref{fig:avmagind} for the spectroscopic sample), the right column is the ``model dependent" magnitude fluctuation described by Eq.~\ref{eq:df} (as in Fig.~\ref{fig:avmag} for the spectroscopic sample).
\label{fig:avmagpz}
} 
\end{figure*}

We use the SDSS spectroscopic and photometric LRG samples, described in \citet{Eis01} and \citet{Coll06}.  We select a rectangular area of the SDSS survey bounded by: $125^{\circ}<\mbox{R.A.}<245^{\circ}$ and $8^{\circ}<\mbox{Dec.}<59^{\circ}$, and covering the redshift range: $0.08<z<0.40$ for the spectroscopic sample, and $0.4<z<0.7$ for the photometric sample. The area was chosen to be contiguous and to be about a degree or so from the survey edges to ensure a uniform as possible coverage. We use the $r$ band model magnitudes and we further filter the catalog by requiring that all the galaxies have a signal-to-noise ratio $\ge2$ and the Galactic extinction in the $r$ band $x_r<0.1$.  
Redshifts are corrected to the local group frame.

To obtain enough signal we must average over enough LRGs in order to beat down the effect of their intrinsic variability in luminosity.  The BF direction found by \cite{WatFelHud09,FelWatHud10} is towards R.A $=180^{\circ}$, Dec. $=-52^{\circ}$. Since the R.A. direction lies within the survey area, whereas the SDSS LRG Dec. distribution are more than $60^\circ$ away, we choose to average the LRG magnitudes across declinations.  For each redshift and R.A. bin we find the average difference in flux units, or ``maggies" (a linear measure of flux defined as $10^{-0.4*m}$), between the magnitudes in that bin and the mean magnitude at any angular position within the same redshift bin:
\begin{eqnarray}
\label{eq:find}
df&=& 10^{-0.4(m_r - \left<m_r\right>_j)}\\
\label{eq:find2}
\delta_m^o &=&<df>_{ij}
\end{eqnarray}
where $i$ refers to the R.A. bin, and $j$ to the redshift bin, $df$ is the difference in LRG flux/magnitude in maggie units and $\delta_m^o$ is the mean value of $df$ in the R.A.-redshift bin $ij$ .  Fig.~\ref{fig:avmagind} shows this magnitude fluctuation in each redshift bin for the spectroscopic sample; the dotted lines show the fluctuation when the magnitudes have not been corrected for Galactic extinction. The left hand side of Fig.~\ref{fig:avmagpz} shows the same for the photometric sample.  

Computing the observed fluctuation in this way is not optimal because we do not have a full sky area with which to compute $\left<m_r\right>_j$, and the result will also be sensitive to the redshift distribution of the LRG within each R.A. bin.  We would also like to compute the observed magnitude fluctuation in such a way that it could be analysed in terms of the theoretical perturbation of observed magnitudes due to peculiar velocities.  Therefore because of these two issues we also calculate the observed fluctuation in R.A., redshift bins in the following way:
\begin{eqnarray}
\label{eq:df}
df&=& 10^{-0.4(m_r - \bar{m}_r)}\\
\label{eq:df2}
\delta_m^o &=&<df>_{ij}
\end{eqnarray}
where again the observed fluctuation $\delta_m^o$ is in flux rather than magnitude units. Eq.\ref{eq:find} and Eq.~\ref{eq:df} differ in their use of $\bar{m}_r$ or $\left<m_r\right>_j$. $\bar{m}_r$ are the expected apparent magnitudes for each LRG in our catalog, assuming a cosmology, no flows (homogeneous universe) and that all LRGs belong to a population with the same mean absolute magnitude and spectral energy distribution (SED), whereas $\left<m_r\right>_j$ are simply the average LRG magnitude (over all angular positions) in redshift bin $j$.   

Fig.~\ref{fig:avmag} shows this fluctuation in the spectroscopic sample for each redshift bin, and with and without Galactic dust corrections applied to their magnitudes, and the right hand side of Fig.~\ref{fig:avmagpz} shows the same for the photometric sample.  

The method we use to calculate $\bar{m}_r$, the expected apparent magnitudes in the absence of any flows for each LRG in our catalog, is outlined below:
\begin{equation}
\label{eq:magth}
\bar{m}_r(z)  = 5 \log_{10}\bar{d}_L(z)+25+\left<M_r\right>_{lrg}+K_r
\end{equation}  
where $\bar{m}_r$ and $\bar{d}_L$ denote the magnitude and luminosity distance in a homogeneous universe respectively, $\left<M_r\right>_{lrg}$ is the average LRG absolute magnitude in the $r$ band, $K_r$ is the $k$-correction, $z$ is the redshift in our filtered LRG catalog.  $\bar{d}_L(z)$ is calculated using the usual equations as given in \citet{hog99}.  The $k$-correction is found by assuming that all the LRGs have an SED with the same shape. It is calculated for each galaxy from:
\begin{equation}
K_{r}=-2.5\log_{10}\left(\frac{1}{(1+z)}\frac{\int d\lambda \lambda f_\lambda(\lambda/(1+z))r(\lambda)}{\int d\lambda \lambda f_\lambda(\lambda)r(\lambda)}\right)
\end{equation}
where $f_\lambda(\lambda)$ is the assumed SED of the LRGs, in this case it is the elliptical SED in \citet{ColWuWee80}, and $r(\lambda)$ is the SDSS $r$ filter.

For calculating the mean absolute magnitude $\left<M_r\right>_{lrg}$ we use the SDSS LRG luminosity function (LF) measured by \citet{WakNicEis06} corrected to redshift zero. To calculate $\left<M_r\right>_{lrg}$ we simply integrate the LF, $\phi(M)$ multiplied by $M_r$.
\begin{equation}
\left<M_r\right>_{lrg}= \frac{ \int \phi(M) M dM}{ \int \phi(M) dM}
\end{equation}
From the above LF we find that we need to average over around 1000 galaxies in each bin so we can be confident that the average absolute magnitude of LRGs in each bin $i$, $\left<M_r\right>_i\equiv \left<M_r\right>_{lrg}$.

Finally this ``theoretical homogeneous universe" LRG catalog is subtracted from the real LRG catalog, to get $df$ as defined in Eq.~\ref{eq:df}.

We now discuss the assumptions made in Eq.~\ref{eq:magth} when calculating the theoretical catalog.  First of all we assume a fiducial set of cosmological parameters to calculate the luminosity distance $\bar{d}_L$.  The choice of parameters has a predictable effect on the resulting observed fluctuation $\delta_m^o$, shifting the whole flucutation upwards or downwards slightly, i.e. changing the normalisation, but it cannot produce an effect which varies with R.A..  Secondly we assume the value we calculate for $\left<M_r\right>_{lrg}$ is in fact the true average LRG absolute magnitude in the bin.  As long as there are over 500 galaxies in each bin (and we have arranged the binning such that this true) and taking into account the errors on the luminosity function from \citet{WakNicEis06}, the error on this quantity is less than 1\%, smaller than the SDSS photometric errors.  There should be no reason to expect  $\left<M_r\right>_{lrg}$ to vary with position, however it could evolve with redshift.   Finally to compute the $k$ correction we must assume the elliptical SED in \citet{ColWuWee80} is truly representative of the actual LRG SED.  Again we find no reason that this assumption, if wrong, would cause an effect where $<\bar{m}_r>$ varied with R.A.. In fact using a completely different SED, the spiral galaxy Sbc from \citet{ColWuWee80}, causes a negligible difference in the results because the difference in $k$-correction values between these two SEDs are small.

It seems that all the assumptions made in calculating this theoretical LRG catalog could cause the same systematic effect: a change in the normalisation of the plots.  This normalisation could be a function of redshift: the difference between $\bar{d}_L$ calculated with different cosmological parameters increases as a function of $z$, and $\left<M_r\right>_{lrg}$ could evolve with $z$.  Crucially, however, none of \textit{these} systematics are expected to cause a change in $\left<m_r\right>$ with angular position.

In this paper we want to constrain the parameters of the BF: $v_b$, $\alpha_{b}$, $\delta_{b}$. The observed fluctuation as defined by Eq.~\ref{eq:df} is the one which is compatible with how the predicted magnitude fluctuations are modelled, as described in Sec.~\ref{sec:theory}.  Therefore to avoid issues with the ``nuisance" systematics outlined above we divide out the normalisation of the fluctuations.  We do this by dividing out the value: MAX($\delta_m^o$)/2 + MIN($\delta_m^o$)/2 from the fluctuations shown in Fig.~\ref{fig:avmag}.

Looking by eye at Fig.~\ref{fig:avmagind} and ~\ref{fig:avmag} it seems the first redshift bin might be contaminated by LRGs which are not at rest with respect to the CMB, or indeed by objects from the Main SDSS sample.  It is not unexpected that ``normal" flows in the universe could still be affecting the observed magnitudes at these redshifts.  Therefore in Sec.~\ref{sec:res} we repeat our analysis using all redshift bins, removing the lowest redshift bin, and removing the two lowest redshift bins to reduce the possibility of either of these two effects.

\begin{figure}
\begin{tabular}{c}
\includegraphics[width=6.5cm]{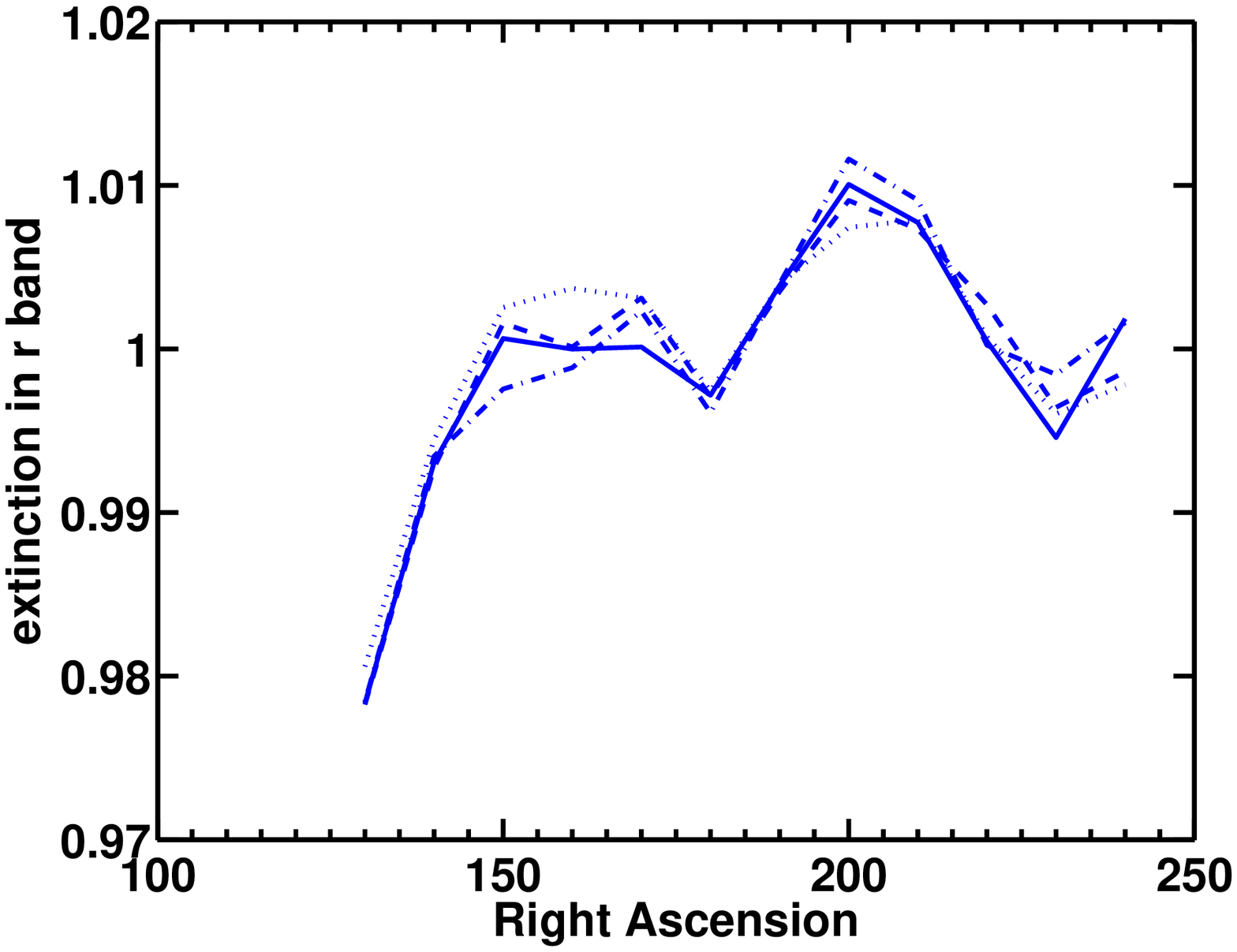}\\
\includegraphics[width=6.5cm]{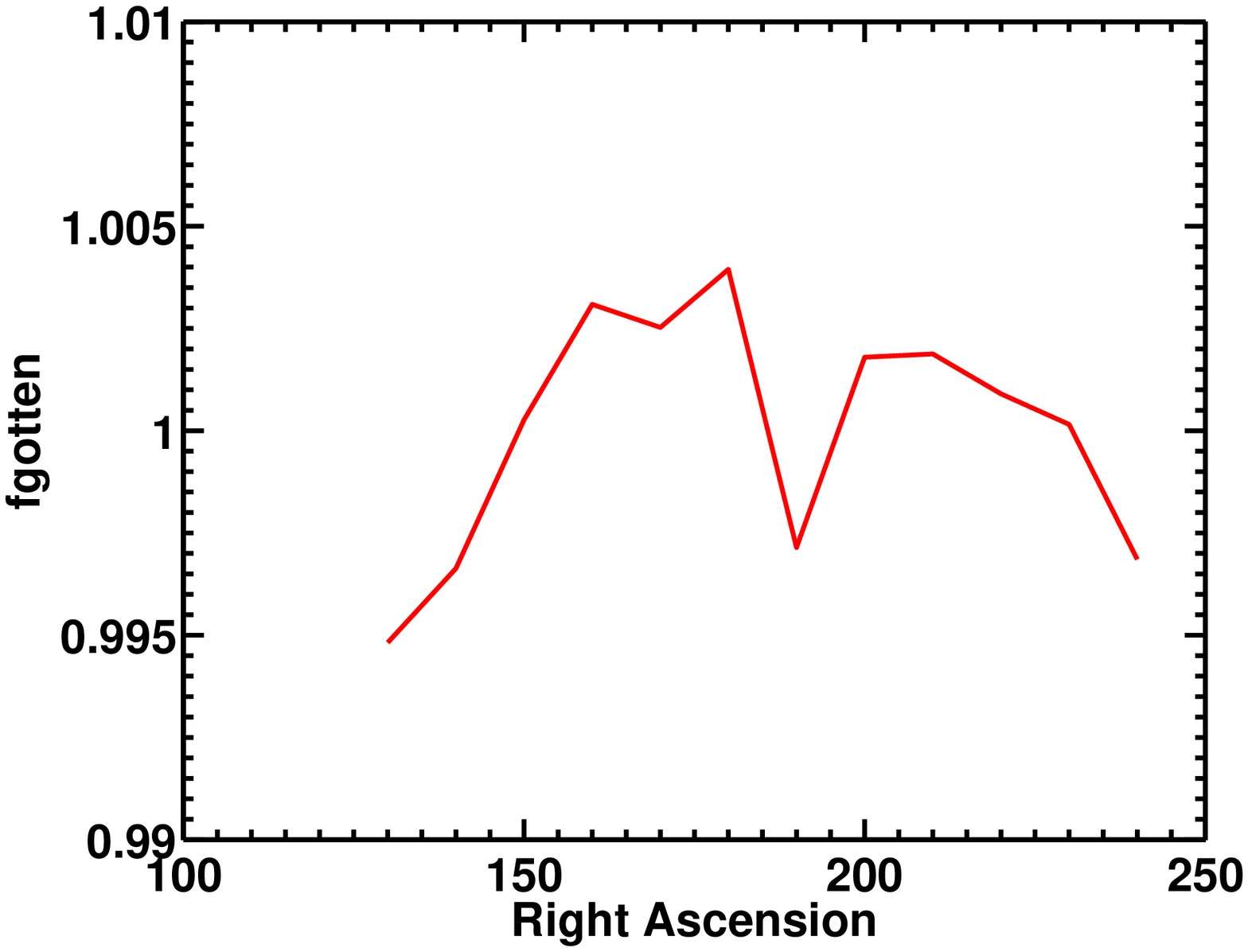}
\end{tabular}
\caption{\textit{Top}: Average Galactic extinction in the $r$ band in each redshift and R.A. bin, the different lines correspond to the different redshift bins.
\textit{Bottom}: Average angular selection (\texttt{\textbf{fgotten}}) as a function of R.A.. The variation in \texttt{\textbf{fgotten}} seems to follow the variation in the Galactic extinction, which is not unexpected e.g. less galaxies are likely to be detected in regions of higher extinction.
\label{fig:xas}
} 
\end{figure}

The two main candidates to explain at least part of the observed fluctuation presented in Figures \ref{fig:avmagind}, \ref{fig:avmag} and \ref{fig:avmagpz}, outside of the context of a 
dipole flow, 
are: reddening by our Galaxy, and the variation in the selection of SDSS LRGs.  These can also be considered to be systematics. It is unlikely that Galactic reddening is a major contributor because i) there is so little difference in the observed fluctuation when using extinction-corrected and non-extinction corrected magnitudes, ii) a similar fluctuation is seen in the $g$, $i$ and $z$ bands i.e. there is no fluctuation in the observed colors.  To see the variation with extinction in the $r$ band, compared to the average extinction, analogously to Eq.~\ref{eq:find} we calculate:
\begin{eqnarray}
\label{eq:dx}
dx&=& 10^{-0.4(x_r - \left<x_r\right>_j)}\\
\delta_x^o &=&<dx>_{ij}
\end{eqnarray}
where $x_r$ is the extinction in the $r$ band at the position of the LRG.  This fluctuation is plotted at the top of Fig.~\ref{fig:xas}.  Each different line style corresponds to a different redshift bin.  Because the Galactic extinction varies only with angular position, and of course not with the redshift of the LRGs, the pattern of mean extinction verses R.A. is the same for each redshift bin: small differences occur only due to the different distribution of LRG angular positions within each redshift bin.

The LRG angular selection function is more difficult to quantify.  To get a handle on what the angular selection function of SDSS LRGs might be we looked at the NYU Value-Added Galaxy Catalog \citep[NYU-VAGC]{nyuvagc}. The NYU-VAGC is a collection of galaxy catalogs cross matched to SDSS designed for the study of galaxy formation, evolution, and large-scale structure.  As such the catalog contains detailed information about the angular selection function.  We choose the parameter
\texttt{\textbf{fgotten}} from the large-scale structure subsample as the value we use to estimate the average LRG completeness as a function of position.  The \texttt{\textbf{fgotten}} parameter describes the fraction of targets in the SDSS Main sample which have successfully measured redshifts. The bottom of Fig.~\ref{fig:xas} plots the mean value of \texttt{\textbf{fgotten}} in each of the R.A. bins divided by the mean value of \texttt{\textbf{fgotten}} over the whole survey area.  Although the variation in \texttt{\textbf{fgotten}} is similar to the fluctuation we observe, its amplitude is significantly smaller.  To incorporate the information contained within the \texttt{\textbf{fgotten}} parameter into our analysis we would also need to know how it varied as a function of flux, and not just angular position.

\section{Theoretical model of magnitude fluctuations}
\label{sec:theory}
The perturbation to the luminosity distance is defined as: $\delta d_L=d_L(z)-\bar{d}_L(z)$, where $d_L(z)$ is the actual observed luminosity distance (i.e. given the universe is inhomogeneous) and $\bar{d}_L(z)$ is the luminosity distance calculated at the \textit{same} observed redshift $z$ but assuming perfect homogeneity.  A bar above quantities indicates that they have been calculated assuming, or are defined by assuming, homogeneity.

The perturbation to the luminosity distance (to first order) due to peculiar velocities is \citep{HuiGree06}:
\begin{equation}
\frac{\delta d_L}{d_L}=\frac{\textbf {v}_e\cdot\hat{\textbf{r}}}{c}-\frac{c(1+z)^2}{\bar{H}\bar{d}_L}\left(\frac{\textbf {v}_e\cdot\hat{\textbf{r}}}{c}-\frac{\textbf {v}_o\cdot\hat{\textbf{r}}}{c}\right).
\end{equation}
where $\textbf {v}_e$ is the peculiar velocity of the emitter, $\textbf {v}_o$ is the peculiar velocity of the observer, $\hat{\textbf{r}}$ is the unit vector in the direction from the observer to the emitter, $z$ is the redshift of the emitter, $\bar{H}$ and $\bar{d}_L$ are the unperturbed Hubble parameter and luminosity distance respectively, calculated from the usual equations as found in \citet{hog99}.

The relation between the luminosity distance and apparent magnitude $m$ is:
\begin{equation}
m=5\log_{10}d_L+M=5\frac{\ln d_L}{\ln 10}+M
\end{equation}
Then differentiating the above equation gives:
\begin{equation}
\delta m=\frac{5}{\ln 10}\frac{\delta d_L}{d_L}
\end{equation}
We assume that the LRG's form a rest frame ($\textbf {v}_e$=0) so we can look at our motion relative to them, our 
dipole 
flow $\textbf {v}_o$, the magnitude fluctuation is now:
\begin{equation}
\label{eq:dm}
\delta m=\frac{5}{\ln 10}\frac{\textbf {v}_o\cdot\hat{\textbf{r}}}{c}\left(\frac{c(1+z)^2}{\bar{H}\bar{d}_L}\right)
\end{equation}
The value of $\textbf {v}_o\cdot\hat{\textbf{r}}$ will depend on the angular position of the LRG (R.A. and Dec., labelled as $\alpha$ and $\delta$ respectively). We model the values by assuming they are due \textit{only} to a BF of magnitude $v_{b}$ and in direction $\alpha_{b}$,$\delta_{b}$, so we can calculate Eq.~\ref{eq:dm} as a function of position ($\alpha$,$\delta$) and redshift $z$.  

\begin{figure}
\begin{tabular}{c}
\includegraphics[width=6.5cm]{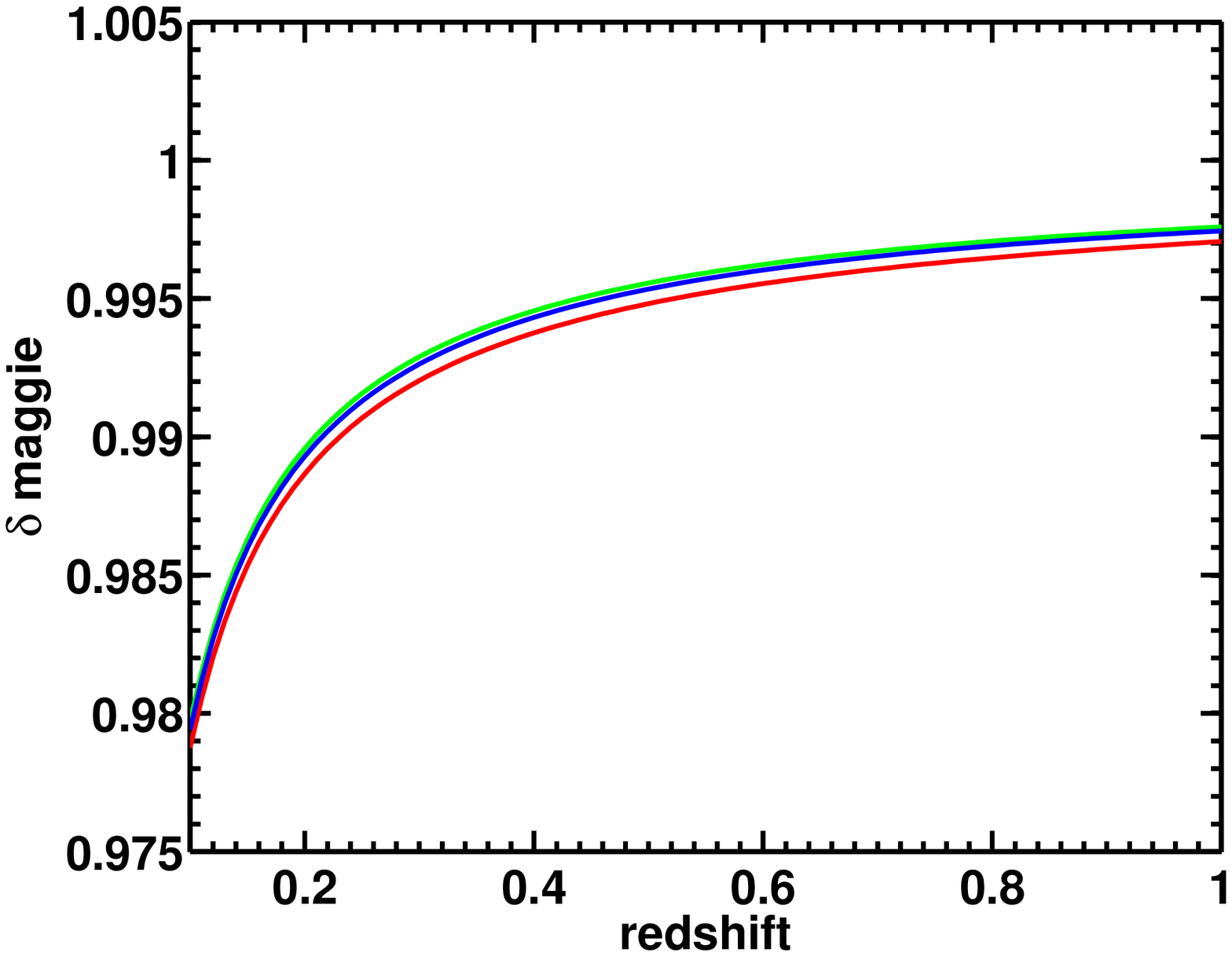}\\
\includegraphics[width=6.5cm]{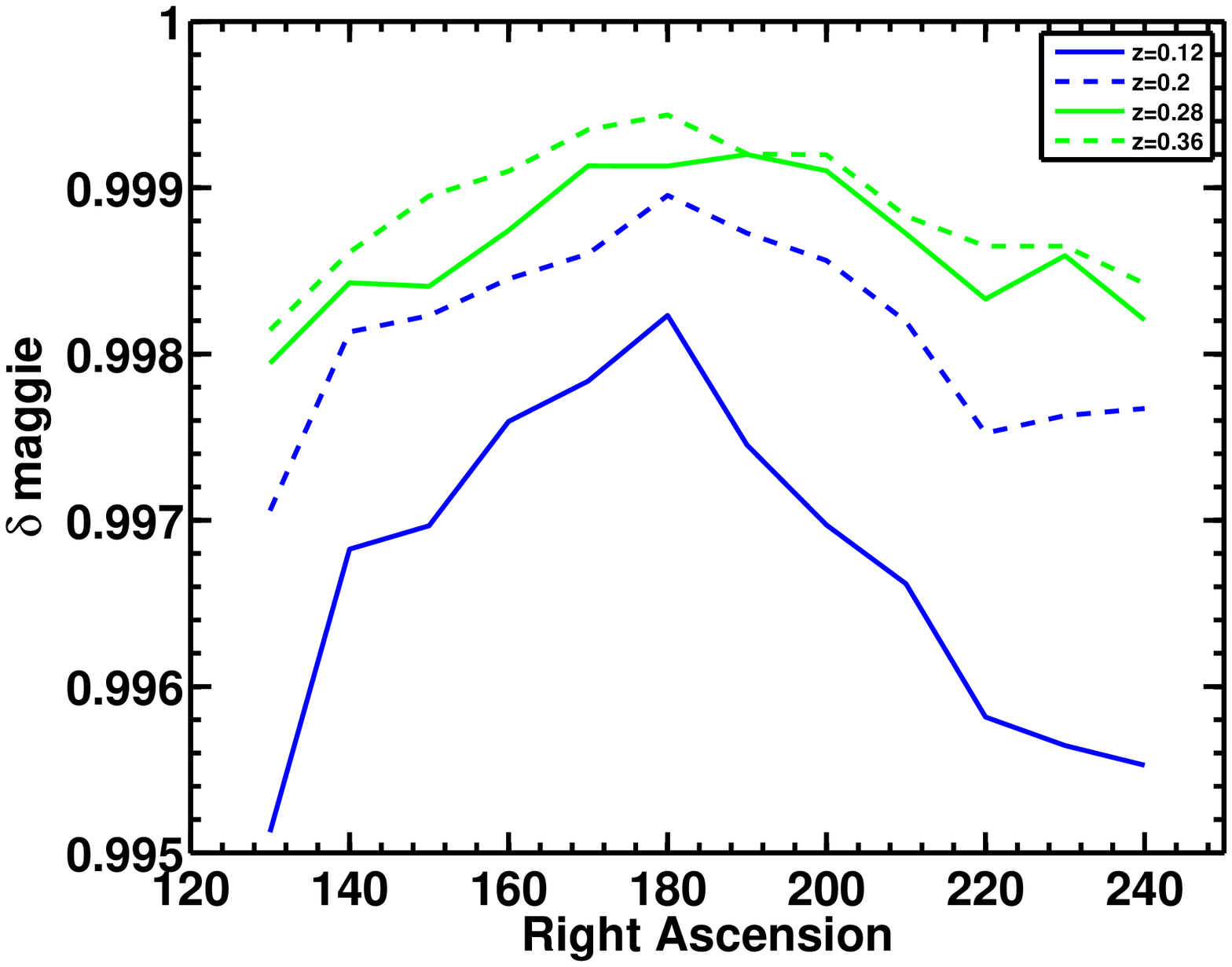}
\end{tabular}
\caption{\textit{Top}: Expected magnitude fluctuation amplitude as a function of redshift due to a 
dipole 
flow with value $v_o\cdot r$ = 300km/s.  Calculated from Eq.~\ref{eq:dm} then transformed into maggies. Three different cosmologies are plotted: $\Lambda$CDM in red; CDM universe ($\Omega_m=1$) in green; Open universe ($\Omega_m=0.3$) in blue. \textit{Bottom}: Expected magnitude fluctuation for our LRG sample in each redshift bin due to a flow with a magnitude of $v_b$ = 300km/s in a $\Lambda$CDM universe.
\label{fig:th}
} 
\end{figure}

The top of Fig.~\ref{fig:th} shows the change in $10^{-0.4\delta m}$ with redshift for a constant value of $\textbf {v}_o\cdot\hat{\textbf{r}} = 300$km/s.  Three different cosmologies are plotted: $\Lambda$CDM in red; CDM universe ($\Omega_m=1$) in green; Open universe ($\Omega_m=0.3$) in blue.  The value of the fluctuation $10^{-0.4\delta m}$ is not a strong function of the cosmological parameters because here (in Eq.~\ref{eq:dm}) they only affect the ``geometric" distance factors. The value of $\textbf {v}_o\cdot\hat{\textbf{r}}$ (and hence the value of the fluctuation) would be primarily driven by the cosmological parameters affecting the growth of structure, in the absence of more exotic explanations.  This is why the choice of cosmological parameters in Eq.~\ref{eq:magth} does not have any significant effect on the ``model dependent" observed magnitude fluctuation. The bottom of Fig.~\ref{fig:th} shows the expected magnitude fluctuation in our LRG sample if there is a 
dipole 
 flow of magnitude $v_b$=300km/s in the direction of $\alpha_b=180$ deg. and $\delta_b$=-50 deg. in a $\Lambda$CDM universe.

\subsection{The expected signal}
 
We expect the magnitude fluctuations generated by a 
dipole 
flow to follow a cosine form (in the magnitudes) across the sky, and we can model this cosine with the following functional form:
\begin{equation}
\delta mag = A\cos(\alpha + B) +C
\end{equation}
where $\alpha$ is the R.A. direction.
Putting this together with Eq.~\ref{eq:dm} we can understand the effect of the flow parameters ($\alpha_b$, $\delta_b$, $v_b$) on the resulting cosine shape of $\delta mag$ vs R.A. in the following way:
\begin{itemize}
\item A is the amplitude and is affected by $\delta_b$ and $v_b$.
\item B is the phase and is affected only by $\alpha_b$.
\item C is the normalisation and is affected by $\delta_b$ and $v_b$, but also the systematics.
\end{itemize}
There is a degeneracy between $\delta_b$ and $v_b$ which both affect the amplitude of the cosine. It also should be noted that two flows with a relation between their bulk angles as: $\alpha_{b1}=\alpha_{b2}+\pi$ and $\delta_{b1}=-\delta_{b2}$ are indistinguishable. Hence we only consider flows with $\delta_b<0$.  This qualitative explanation of the signal also illustrates why we can divide out the constant $C$: it is the only part affected by the systematics, and information on the flow amplitude and declination direction is still contained within $A$.

\section{Fitting the flow}
\label{sec:fit}

\subsection{Errors on the measured $\delta_m$}

Naturally, even in the absence of measurement errors and peculiar velocities, the LRG magnitudes will have some distribution within the redshift bin. Then in the limit where the redshift interval of the bin is infinitesimally small, and all LRGs have exactly the same SED and negligible reddening, then this distribution should follow the LRG luminosity function. If the individual magnitudes are modelled as having a Gaussian measurement error distribution with a width of $\sigma$, and if instead the LF was a $\delta$ function at one magnitude, then the observed distribution of magnitudes in the bin would be a Gaussian of width $\sigma$ around this value.  Therefore we can approximate the $\sigma$ of the error distribution by looking at the difference between the observed distribution of magnitudes in \textit{each bin} and the expected distribution:
\begin{equation}
\label{eq:std}
\sigma_{\delta_m}^2 =\frac{\left< \left(X - \bar{X}\right)^2  \right>}{\sqrt{N}}
\end{equation}
where $X=m_r-\bar{m}_r$: the observed magnitudes minus the expected magnitudes, $\bar{X}$ is the mean of this distribution, and $N$ is the number of LRGs in the bin.

\subsection{$\chi^2$ fitting}
\label{sec:res}

We calculate the following $\chi^2$ statistic:
\begin{equation}
\chi^2(v_{b}, \alpha_{b}, \delta_{b})=\sum_{ij}\left(\frac{\delta_m^o-\delta_m^p(v_{mag}, \alpha_{b}, \delta_{b})}{\sigma_{\delta_m}}\right)^2
\end{equation}
where the sum is over the redshift bins $i$ and the R.A. bins $j$; $\delta_m^o$ is the observed magnitude fluctuation in bin $i,j$ (see Section.~\ref{sec:data}) and $\delta_m^p$ is the predicted value of the magnitude fluctuation in bin $i,j$; $\sigma_{\delta_m}$ is the error on the observed magnitude fluctuation in bin $i,j$ found from Eq.~\ref{eq:std}. 

To calculate the predicted fluctuation, $\delta_m^p$, we calculate the $\delta m^N$ using Eq.~\ref{eq:dm} (given the trial flow parameters) for each of the $N$ LRGs in our catalog, and then average them in the same way as the data: 
\begin{equation}
\delta_m^p = <10^{-0.4 \delta m^N}>_{ij}
\end{equation}

\begin{figure*}
\begin{tabular}{ccc}
\includegraphics[width= 5.5cm]{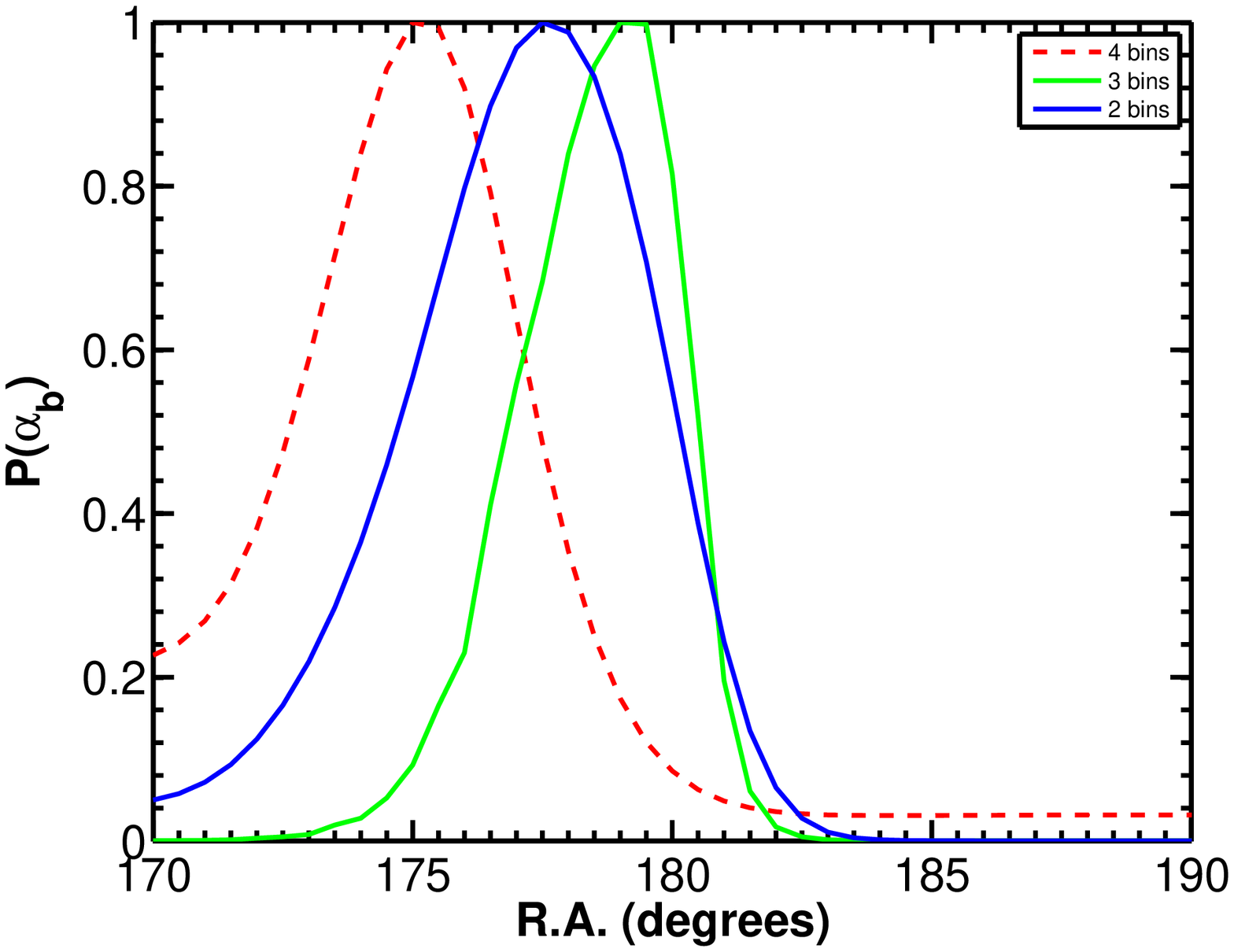}&
\includegraphics[width= 5.5cm]{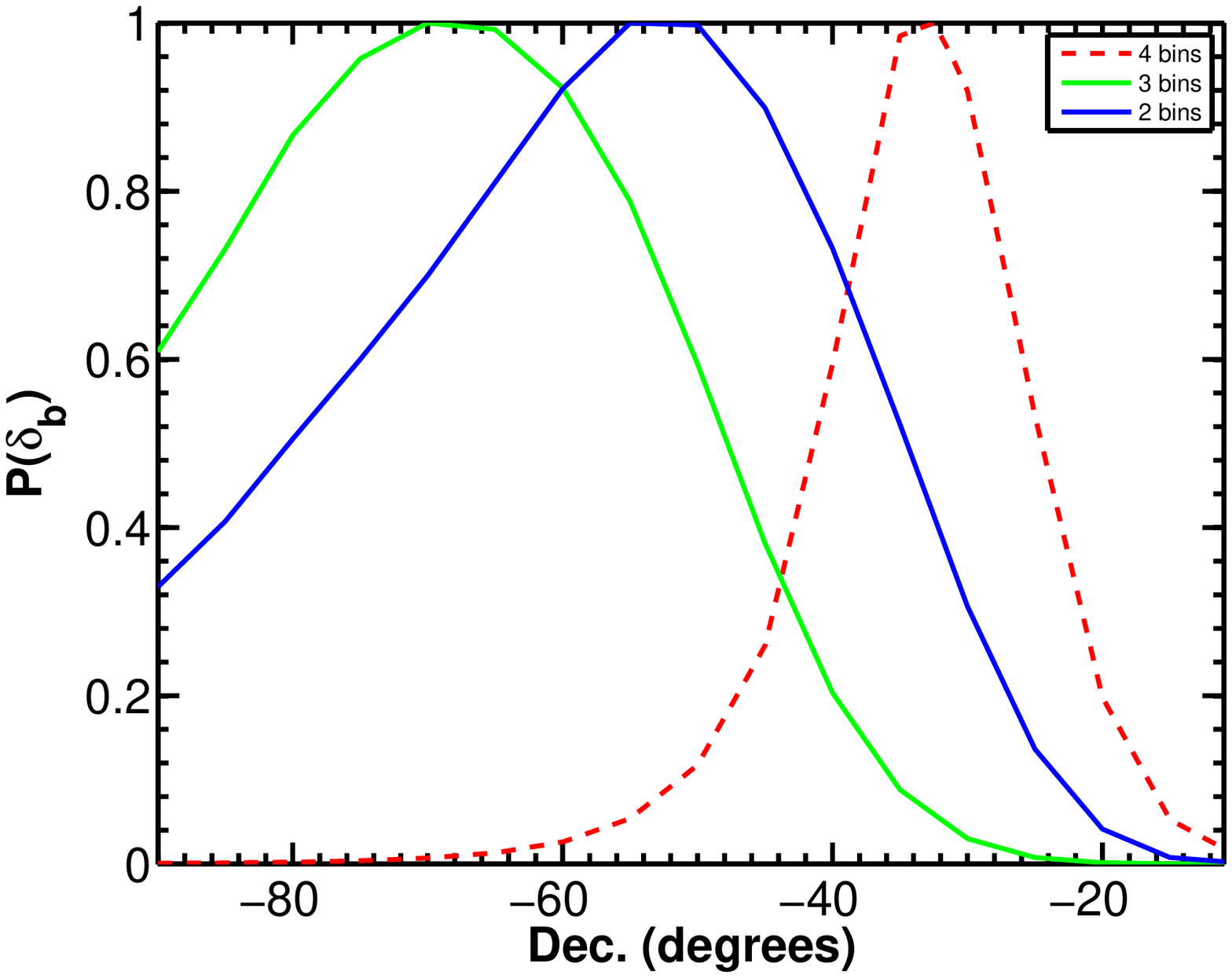}&
\includegraphics[width= 5.5cm]{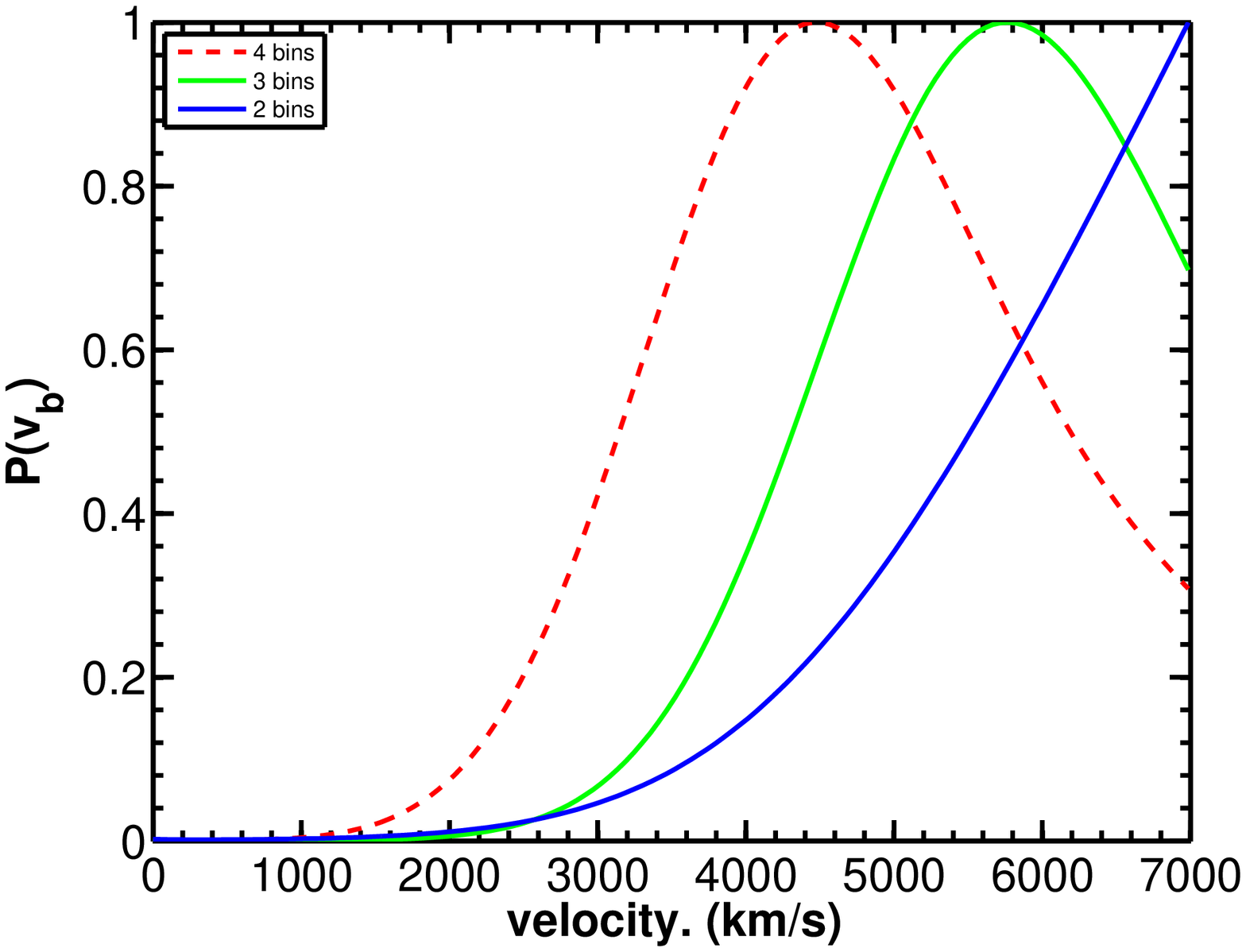}\\
\end{tabular}
\caption{One-dimensional likelihoods of the flow parameters given by $P\propto \exp(-\chi^2/2)$, after marginalising over the other 2 parameters for the spectroscopic catalog.  The red dashed lines correspond to when all 4 redshift bins are used in the analysis; solid green lines when only 3 bins are used; solid blue when only 2 bins are used.
\label{fig:like}}
\end{figure*}

The probability as a function of the flow parameters is calculated from the $\chi^2$ as follows:
\begin{equation}
P(v_{b}, \alpha_{b}, \delta_{b}) \propto \exp\left(-\frac{\chi^2}{2}\right)
\end{equation}
The resulting one-dimensional likelihoods of the flow parameters (after marginalising over the other 2 parameters) are shown in Fig.~\ref{fig:like}. The 1-$\sigma$ constraints on each flow parameter (after marginalising over the other parameters) are shown in Table~\ref{tab:res} for the spectroscopic catalog.

\begin{figure*}
\begin{tabular}{cc}
\includegraphics[width=6.5cm]{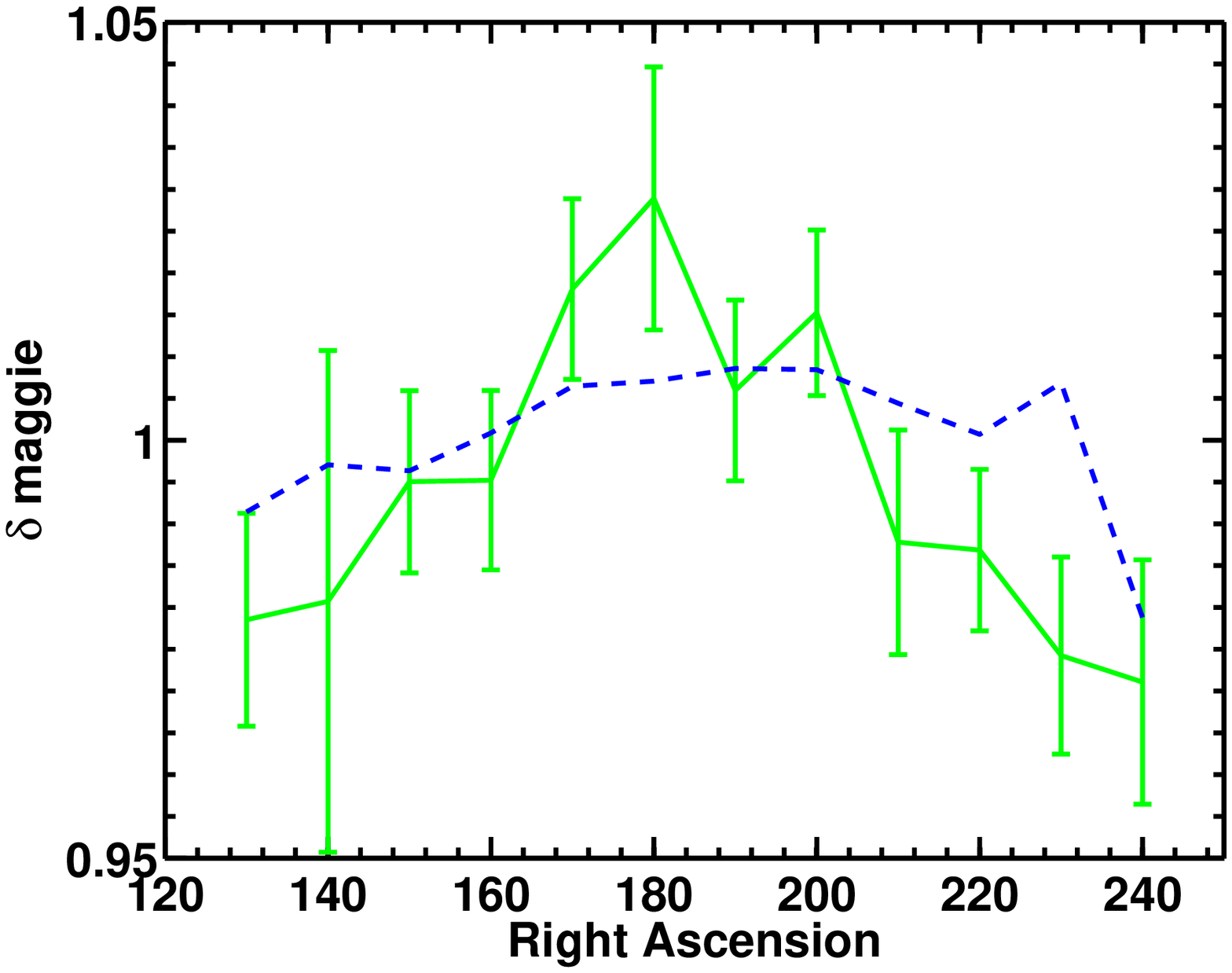}  &
\includegraphics[width=6.5cm]{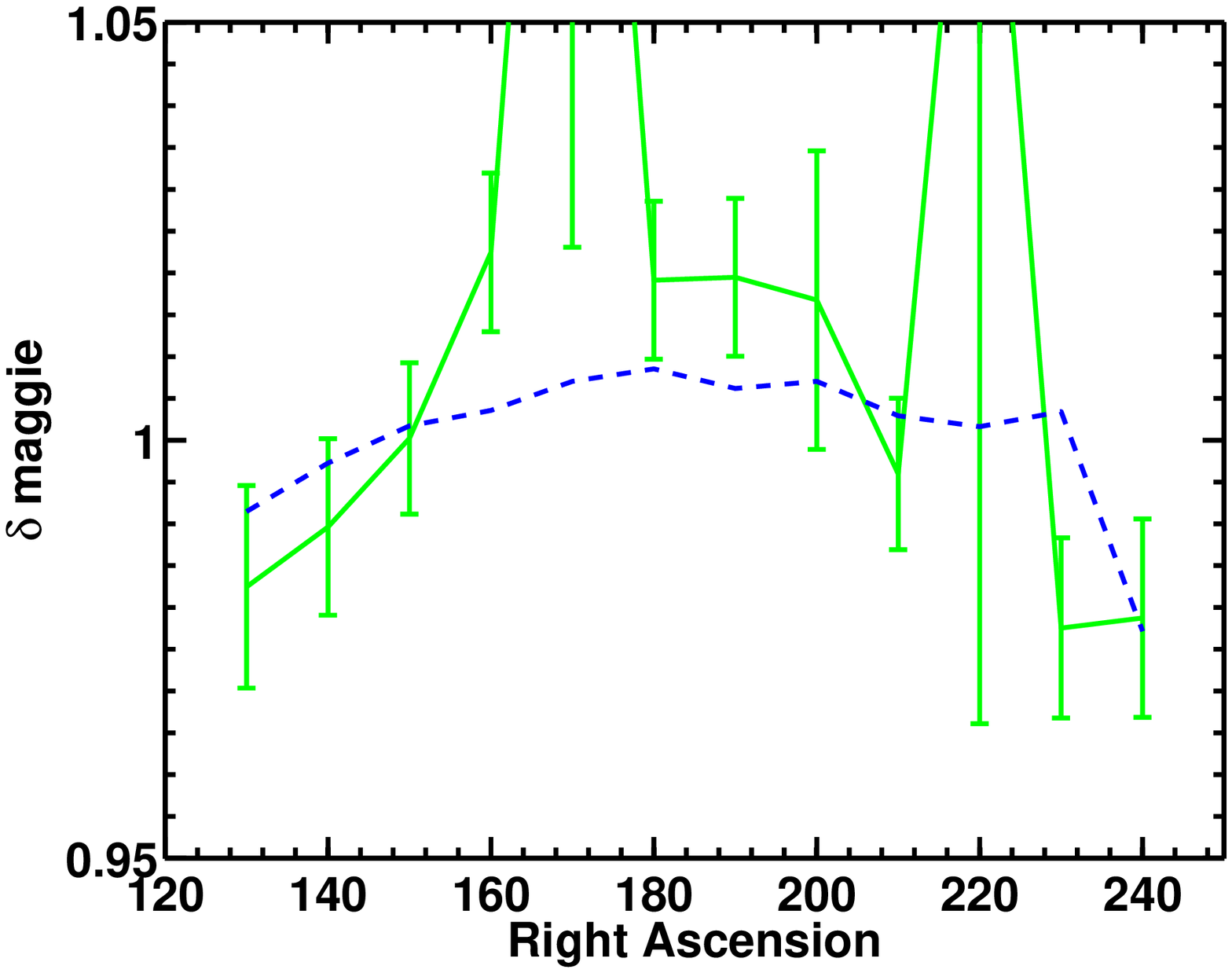} \\
\end{tabular}
\caption{Examples of the 4-redshift bin best-fit model (bottom row of Table~\ref{tab:res}, blue dashed lines) compared to to the data (green solid lines).  The left-hand side shows the $0.24<z<0.32$ bin and the right-hand side shows the $0.32<z<0.40$ bin. \label{fig:fit}} 
\end{figure*}

Fig.~\ref{fig:fit} shows the magnitude fluctuation data in two highest redshift bins with the best-fit model from Table~\ref{tab:res} over-plotted.  The best-fit flow model was taken from the fit including all four redshift bins.

\subsection{Direction fitting}
\label{sec:dir}

We also attempted a separate approach to find the direction of the flow by fitting a vector of the directions of the LRG's weighted by each galaxy's magnitude (or maggie). We find the weighted mean R.A. of the fluctuations by weighing the R.A. of each LRG with the magnitude $m^n_r$ or with its "maggie" $10^{-0.4(m^n_r-<m_r^n>)}$ where $m^n_r$ is the r-band magnitude of each LRG indexed by $n$ and $<m_r^n>$ is the mean of the distribution. We have done that for the whole distribution (all redshifts) and separately by looking at the same redshift regions as described in Fig.~\ref{fig:avmagind}, we also calculated the R.A. direction in various Dec. regions. The results are not very sensitive to the binning schemes and agree by and large with the analysis above. We found that the most likely R.A. is $178\pm11^\circ$ in good agreement with the binned analysis presented here and with the velocity analyses presented in \citet{WatFelHud09,FelWatHud10,MaGorFel11,KasAtrKoc08,KasAtrKoc10}.

\begin{table}
\caption{Results from the spectroscopic catalog.\label{tab:res}}
\begin{center}
\begin{tabular}{|c|c|c|c|}
\hline
No. $z$ bin & $\alpha_b$ & $\delta_b$ & $v_b$\\
\hline
\hline
2 & 177.0$^{+2.7}_{-1.4}$  &-50$^{+12}_{-21}$  & $>$7000\\
\hline
3 & 179.0$^{+1.3}_{-1.7}$   &-65$^{+10}_{-18}$ & 6000$^{+1000}_{-900}$\\
\hline
4 & 185.5$^{+1.9}_{-3.0}$   &-35$^{+10}_{-6}$  & 4450$^{+1350}_{-1100}$\\
\hline
\end{tabular}
\end{center}
\end{table}

\section{Discussion}
\label{sec:disc}


Our goal in this paper was to search for a signature of a bulk flow by looking for fluctuations in the magnitudes of distant LRGs.  

Figures \ref{fig:avmagind}, \ref{fig:avmag} and \ref{fig:avmagpz} show that we do find a coherent fluctuation in the LRG magnitudes which seems to be fairly independent of their redshift.  The cosine-like shape of the fluctuation matches that which would be expected if this observed fluctuation is due to a 
dipole motion with respect to the LRG sample.
The maximum of the fluctuation is also located roughly at the R.A. direction of a bulk flow found by other authors (170$\pm10^{\circ}$), however its amplitude is more than an order of magnitude larger than what would be expected from Eq.~\ref{eq:dm} after assuming a ``reasonable" flow (Fig.~\ref{fig:th}).  We find a magnitude fluctuation on the order of a few percent in flux for the spectroscopic sample, and about a percent for the photometric sample.  Comparatively one would expect a fluctuation of less than 1 percent if the flow had an amplitude of less than 1000km/s.  Figures \ref{fig:avmagind}, \ref{fig:avmag} and \ref{fig:avmagpz} do also show the amplitude of the observed fluctuations decreases with redshift, as expected from Eq.~\ref{eq:dm} and illustrated by Fig.~\ref{fig:th}.

Our results are extremely sensitive to systematic effects, which is an obvious alternate explanation for the magnitude fluctuations we find.  Possible effects could include: photometric zeropoint variation, correlated errors in the Galactic extinction correction applied to the magnitudes, and selection effects on the LRG sample.  

The SDSS imaging data is acquired in a continuous scan, and each scan is obtained along a stripe. The survey strategy means the data we average together end up coming from different stripes. However, it would be very unlikely that a zeropoint variation could produce such a coherent fluctuation after averaging across different stripes.  

Galactic extinction on the other hand will have an effect which is correlated with the angular position of the line of sight.  Figs.~\ref{fig:avmagind} and \ref{fig:avmag} show that there is very little effect on shape of the observed fluctuation when no Galactic extinction correction is applied to the magnitudes.  This implies that an error in the Galactic extinction estimate is very unlikely to be the sole source of the fluctuation.  In Fig.~\ref{fig:xas} we plotted the average Galactic extinction in each bin.  The shape of the extinction variation is extremely similar to our observed magnitude fluctuation, but its amplitude cannot fully explain the entire signal that we find. 

The LRG angular selection function, however, is likely to have a big effect on our results.  The angular selection varies in a similar manner to the fluctuation we observe.  It would seem from comparing the bottom panel of Fig.~\ref{fig:xas} to figures \ref{fig:avmagind} and \ref{fig:avmag}, that this could only contribute to about 1 percent of the observed fluctuation.  It is not clear however how to take account of this in our analysis, one would need to know the angular selection as a function of LRG magnitude before precise constraints on the magnitude of the flow could be made.

We fitted a flow model to the observed fluctuation, and constrained the three flow parameters, its direction and magnitude: $\alpha_b$, $\delta_b$, $v_b$.  We found that the flow R.A. direction was consistent with the direction found by other authors. The R.A. direction of the flow was the best constrained parameter, unsurprising because it is the only parameter where the rough best-fit value can be inferred from Fig.~\ref{fig:avmagind} and ~\ref{fig:avmag} by eye.  It is also unsurprising, given the amplitude of our signal, that we find an anomalously large flow. We do not attempt to perform the fit using the photometric sample.  Although it has the advantage over the spectroscopic sample in that it has over 6 times as many LRGs, and all of them are at redshifts where we are most confident that they are truly at rest with respect to the CMB, it is non-trivial to account for their redshift errors in the analysis. At this stage given the small size of the survey area and its direction we do not think this is a worthwhile exercise. 

We repeated the fit using the spectroscopic sample leaving out the lowest, and second lowest redshift bins,  since these redshift bins may contain LRGs that are not at rest with respect to the CMB.  As Fig.~\ref{fig:th} shows these are also the bins which should contain the most signal.  We find that when leaving out these bins some of the constraint on the declination direction of the flow is lost but the right ascension results are robust. We also found that the magnitude weighted right ascension directional fit also agrees with our binned analysis well.

Since our results indicate that our local group is moving with respect to the SDSS LRG sample, we infer that the source of our local motion is from scales comparable to the LRG scale, $z=\sl O(0.1-0.2)$. This result agrees with \citet{FelWatHud10} that found that the sources responsible for the bulk flow are at an effective distance of $\gteq 300$ \hmpc, \ie well within the horizon, and contradicting the suggestion of a coherent flow on much larger scales claimed by \citet{KasAtrKoc08, KasAtrKoc10} and modeled as a {\it tilted} Universe \citep{KasAtrKoc08,MaGorFel11}. A note of caution must be made here, the uncertainties and systematic effects on the magnitude of the flow suggested here, precludes anything more than a hint of a subhorizon flow.  We have presented results using only the $r$ band data, using either the $g$ or $i$ band data instead does not significantly alter our conclusions.

\cite{ItYaTa10} used galaxy catalogs (not just the LRGs) constructed from the SDSS Data Release 6 and looked at the variations in the pixelised number counts, a similar approach to that presented in this paper.  They found the probability of $\beta$ ($\approx v_{bulk}/c$) was consistent with a zero bulk flow, but also was not inconsistent with flows with $\beta\simeq0.01$.  In fact, the galaxy sub-samples closest to those used here (termed Northern Galactic Hemisphere - shallow samples) find values of $\beta\geq\sim0.02$, and are only marginally consistent with zero. This indicates that the analysis both presented here and in \cite{ItYaTa10} may be seeing the same effect in the SDSS data: either the signature of a large scale flow or a systematic in the data.  
However, in a similar method to \cite{ItYaTa10}, \cite{BlakeWall02} analyzed the radio galaxy distribution from the NRAO VLA Sky Survey and detected a cosmological dipole anisotropy that is consistent with the CMB dipole in both amplitude and direction.

We also note that a non-negligible detection of a fluctuation in galaxy magnitudes should also show up as a detection of excess large scale power in the clustering of galaxies. \cite{ThoAbdLah11} used a similar data set to the one employed here, the MegaZ DR7 photometric LRG sample, and found excess power on large scales of between 2-4$\sigma$ significance by looking at the angular clustering of the LRGs.

\cite{NusBraDav11} also used galaxy luminosities to study the bulk flow.  Although they found a bulk flow with a similar direction to that found here, and by other authors, they do not find the flow to have an anomalously large magnitude as compared to that expected from $\Lambda$CDM. This contention between different measures of the bulk flow using kSZ, galaxy distances, galaxy luminosities shows that further studies of our local velocity field are vital to test the current paradigm.

Upcoming surveys should be able to confirm or repudiate this detection.  The Large Synoptic Survey Telescope \citep[LSST]{SciBook} is potentially the best candidate for two reasons: its ability to discover of order 50,000 photometric SNIa per year, and its wide area coverage (20,000 sq.deg.).  Most of these SNIa will not be able to be followed up spectroscopically, however it is expected that photometric redshifts of SNIa observed by LSST could reach a precision of just 1\%.  Ideally a multi-wavelength approach, which would necessarily involve different targets and different instruments, would provide the strongest evidence in confirming the existence of this signal. Further, the whole sky data from WISE (Wide-field Infrared Survey Explorer) satellite \citep{WISE10} will become available in the next few years. It will be sensitive to 10\% overdensities out to $\sim 700$\hmpc\ ($z\sim 0.2$) and although it does not measure redshifts, it may be used to search for over-- and under--densities in the direction and anti--direction of the flow.
\\

\noindent{\bf Acknowledgement:} 


HAF was supported in part by an NSF grant AST-0807326 and by the University of Kansas GRF and would like to thank Sarah Bridle for useful conversations.
AA would like to thank Tim Axelrod, Daniel Eisenstein, Mike Hudson, Eyal Kazin, Marc Metchnik and Brian Schmidt for their useful input and suggestions.

\bibliographystyle{mn2e}
\bibliography{haf}

\end{document}